\documentclass[preprint, 12pt]{aastex}
\usepackage{cite}
\usepackage{rotating}
\usepackage{graphicx}

\newcommand{\msun}{M$_{\sun}$}

\newcommand{\ldl}{$\lambda/{\Delta}{\lambda}$}

\newcommand{\meth}{CH$_4$}

\newcommand{\kms}{km~s$^{-1}$}

\begin{document}

\title{A Search for Photometric Variability in L and T type Brown Dwarf Atmospheres}
\author{Harish Khandrika\altaffilmark{1},
 Adam J.\ Burgasser\altaffilmark{1}, Carl Melis\altaffilmark{1},
Christopher Luk\altaffilmark{1}, Emily Bowsher\altaffilmark{1,2}, 
 and Brandon Swift\altaffilmark{3}}

\altaffiltext{1}{Center for Astrophysics and Space Sciences, University of California, San Diego, CASS, M/C 0424
9500 Gilman Drive, La Jolla, CA 92093-0424; hkhandrika@alumni.ls.berkeley.edu}
\altaffiltext{2}{Department of Astronomy, Columbia University, Mail Code 5246, 550 West 120th Street, New York, New York 10027}
\altaffiltext{3}{Department of Astronomy/Steward Observatory University of Arizona, 933 North Cherry Avenue, Rm. N204, Tucson, AZ 85721-0065}

\begin{abstract}
Using the Gemini infrared camera on the 3-meter Shane telescope at Lick Observatory, we have searched for broad-band $J$ and $K^{\prime}$ photometric variability for a sample of 15 L and T-type brown dwarfs, including 7 suspected spectral binaries.   Four of the dwarfs--- 2MASS~J0939$-$2448, 2MASS~J1416+1348A, 2MASS~J1711+2232, and 2MASS~J2139+0220---exhibit statistically significant variations over timescales ranging from $\sim$0.5~hr to 6~days. Our detection of variability in 2MASS~J2139+0220 confirms that reported by Radigan et al., and periodogram and phase dispersion minimization analysis also confirms a variability period of approximately 7.6$\pm$0.2 hours. Remarkably, two of the four variables are known or candidate binary systems, including 2MASS~J2139+0220, for which we find only marginal evidence of radial velocity variation over the course of a year. This result suggests that some spectral binary candidates may appear as such due to the blending of cloudy and non-cloudy regions in a single ``patchy'' atmosphere. Our results are consistent with an overall variability fraction of $35\pm5\%$, with no clear evidence of greater variability among brown dwarfs at the L dwarf/T dwarf transition.
\end{abstract}

\section{Introduction}
Low-temperature, low-mass stars and brown dwarfs of spectral types L and T exhibit a broad diversity in color and spectral morphology arising from variations in effective temperature (T$_{eff}$), surface gravity, composition, atmospheric dynamics and cloud properties  (\citealt{2005ARA&A..43..195K} and references therein). Clouds are a particularly unique and important component of L and (some) T dwarf atmospheres, where the photospheric conditions are conducive to the formation of mineral and metal condensates \citep{1996A&A...305L...1T,2001A&A...376..194H,2002ApJ...577..974L}. Evidence for condensate formation is found in pronounced changes in spectral features (e.g., the depletion of TiO and VO through the L dwarf sequence; \citealt{2000AJ....120..447K}), red near-infrared colors ($J-K \approx$ 1.5--2.5; \citealt{2000ApJ...542..464C,2001ApJ...548..908L,2004AJ....127.3553K}) and the direct detection of silicate grain absorption at mid-infrared wavelengths \citep{2008ApJ...678.1372C, 2008ApJ...674..451B, 2008ApJ...686..528L}. While one-dimensional atmospheric models have had modest success in reproducing some of the spectral and color characteristics of cloudy L dwarfs
(e.g., \citealt{2001ApJ...556..872A,2003ApJ...586.1320C,2005ApJ...621.1033T,2008MNRAS.391.1854H}), they provide no insight into the distribution of cloud structures across the surfaces of a rotating brown dwarfs (i.e., the presence of banding, vortices and jets), nor can they predict the long-term evolution of these structures. 
Cloud effects are also important for young and/or closely-orbiting hot exoplanets, 
which have comparable equilibrium temperatures and evidence of unusually thick 
photospheric clouds \citep{2010ApJ...725.1405B,2011ApJ...729..128C,2011ApJ...737...34M, 2011ApJ...733...65B}

While the relatively small sizes ($ R \sim R_{Jupiter}$) and large distances of brown dwarfs prohibits resolved imaging of cloud features, photometric variability proves to be an effective indirect probe (see recent reviews by \citealt{2005AN....326.1059G} and \citealt{2005ESASP.560..429B}). Studies over the last decade have shown that perhaps as many as 70\% of L dwarfs are variable in broad-band photometry (e.g., \citealt{2006A&A...448.1111R}), many of which are too cool to exhibit magnetic spotting \citep{2002ApJ...577..433G, 2002ApJ...571..469M}.
The variations are typically
periodic over timescales of $\sim$2--10 hours, consistent with rotation periods \citep{2004A&A...419..703B,2008ApJ...684.1390R}. There is also evidence of changes in light curve behavior over multiple rotation cycles \citep{2001A&A...367..218B,2009ApJ...701.1534A}. Variability appears to be more common for dwarfs at the transition between the L and T dwarf classes (L8 to T5), sources which also exhibit a remarkable brightening between 1--1.3$\mu$m \citep{2003AJ....126..975T, 2006ApJS..166..585B, 2006ApJ...647.1393L,2008ApJ...685.1183L}. Possible origins of both the variability and brightening include a sudden rain-out of condensates from the photosphere \citep{2004AJ....127.3553K} or fragmenting of the cloud layer \citep{2001ApJ...556..872A, 2002ApJ...571L.151B}. 

Unfortunately, the majority of brown dwarf variable detections have small amplitudes ($\lesssim$0.1~mag), and few sources exhibit truly significant light curves \citep{2009ApJ...701.1534A, 2012ApJ...750..105R}. This makes it difficult to explore cloud meteorology in a statistically robust manner, or quantify the dependence of variability on other parameters such as rotation, temperature, surface gravity, metallicity, or overall cloud content.
In an effort to increase the sample of variable dwarfs and more fully explore brown dwarf cloud morphologies, we have conducted a near-infrared variability study of 15 L and T-type dwarfs using the Gemini near-infrared camera at Lick Observatory.  In Section 2 we describe our observation methods, data reduction procedures and photometric analysis. In Section 3 we present the light curves for all of our sources, and describe the statistical analyses used to identify significant variability and search for periodicities. We discuss individual variable sources in detail in Section 4, where we also present additional spectroscopic observations of the previously-identified variable dwarf 2MASS~J2139+0220\footnote{Throughout the text, we shorten source names to 2MASS/SDSS J$hhmm{\pm}ddmm$, were $hhmm$ and ${\pm}ddmm$ are the Right Ascension (hours and minutes) and declination (degrees and arcminutes), respectively, at equinox J2000.  Full source coordinates are given in Table~\ref{sourcetable}.} \citep{2012ApJ...750..105R}.   Results are discussed in Section 5 and summarized in Section 6.

\section{Observations}

\subsection{Sample Selection}

We selected sources for our variability study using the online DwarfArchives catalog\footnote{\url{http://DwarfArchives.org}.} of over 800 known L and T dwarfs. We narrowed this sample to sources visible in the northern hemisphere ($\delta > -20\degr$), and focused on two categories of targets.  First, we selected L and T dwarfs with  $J < 15.5$ and $K_s < 15$, whose 2MASS $J-K_s$ colors are $>$0.25 mag redder or bluer than the mean color for their spectral class as quantified in \citet{2009AJ....137....1F}.  Such sources may have with unusually thick or thin clouds, respectively (e.g., \citealt{2004AJ....127.3553K,2008ApJ...674..451B, 2008ApJ...686..528L}), allowing us to examine variability dependence on total cloud content.
We excluded sources spectroscopically known to be young/low-gravity objects, which are red due to reduced collision-induced H$_2$ opacity (e.g, \citealt{2006ApJ...639.1120K,2007ApJ...657..511A}); and metal-poor subdwarfs, which are blue due to enhanced H$_2$ opacity \citep{1994ApJ...424..333S,2003ApJ...592.1186B}. 
We required at least two additional, comparably bright stars be present within Gemini's field of view (FOV).
Our second sample comprised spectral binary candidates, sources whose spectra appear to be a blend of two distinct spectral types \citep{2007AJ....134.1330B}, with the goal of measuring potential eclipse events by these unresolved systems. 
For these sources we relaxed the declination, brightness and color constraints, but not the requirement for at least two bright calibrator stars.
The final list of targets, which includes four red dwarfs, two blue dwarfs and nine spectral binary candidates, is provided in Table~\ref{sourcetable}.

\subsection{Lick/Gemini Near-Infrared Imaging}

Observations were undertaken at the Lick Observatory 3m Shane Telescope with the Gemini dual-channel infrared camera \citep{1993SPIE.1946..513M}. The Gemini camera allows simultaneous imaging in two near-infrared bands over a  2$\farcm9\times2\farcm9$ FOV with resolution 0$\farcs$68~pixel$^{-1}$, permitting measurement of both brightness and color variability.  We used the broad-band $J$ ($\lambda_c$ = 1.25 $\mu$m) and $K^{\prime}$ ($\lambda_c$ = 2.12 $\mu$m) bands.  The former samples a region with low molecular gas opacity, and is thus more sensitive to variations in cloud opacity \citep{2001ApJ...556..872A}; the latter samples a region dominated by collision-induced H$_2$ absorption. 

Data were obtained during two runs,  2010 March 23-29 and 2010 October 25 -November 1 (UT); each set of observations will hereby be referred to using their respective months (i.e. March and October). Observations are summarized in Table \ref{observationtable}.
For the March observations, we performed continuous monitoring over 1--2 hours, obtaining 45s exposures with a repeated 4-position dithering sequence. For the October observations, we obtained single sets of 5-position, 30s dithered exposures interleaved over 2 to 4 hour periods. For each revisit, care was taken in repositioning the source on the same pixel at the start of each dither sequence to minimize pixel sensitivity variations. 

Data were reduced using custom routines developed in the IDL\footnote{Interactive Data Language} environment.  Raw image frames were first corrected for instrumental and telescope response by dividing through by a normalized flat-field frame, constructed from median-combined observations of the twilight sky typically taken on the same night or one night prior/afterwards. Flat-fielded images were dark-subtracted and bad pixels replaced with nearest neighbors. The dithered science images were then pair-wise subtracted, manually taking into account source overlap in the selection of subtraction pairs.  Analysis (Section~3.1) was based on these reduced images.

\subsection{Keck/NIRSPEC Near-Infrared Spectroscopy}

One of our targets, J2139+0220, was identified by \citet{2010ApJ...710.1142B} as a spectral binary candidate based on its unusually strong {\meth} absorption at 1.2~{\micron}, and inferred component types of L8.5$\pm$0.7 and T4.5$\pm$1.5. \citet{2012ApJ...750..105R} report that the source is unresolved in
{\em Hubble Space Telescope} Near-Infrared Camera and Multi-Object Spectrometer (NICMOS; \citealt{1992SSRv...61...69T}) observations, ruling out the presence of binary companion wider than $\sim$1~AU.  To examine whether J2139+0220 is simply a tighter binary, we obtained three epochs of high resolution near-infrared spectroscopy of this source
using the NIRSPEC echelle spectrograph on the Keck II telescope \citep{1998SPIE.3354..566M}.
Data were obtained on 2010 November 26, 2011 July 6 and 2011 September 7 (UT); in all cases conditions were clear to light cirrus, with seeing of 0$\farcs$7--1$\arcsec$ at $J$-band.
We used the N7 filter and 0$\farcs$432$\times$12$\arcsec$ slit to obtain 2.00--2.39~$\micron$ spectra over orders 32--38 with {\ldl} = 20,000 ($\Delta{v}$ = 15~{\kms}) and dispersion of 0.315~{\AA}~pixel$^{-1}$.  An AB nodding sequence was obtained with individual exposure times of 1000~s each.  We also observed the nearby A0~V star HD~198070 (V = 6.37) for flux calibration and telluric correction, and dark, quartz lamp and NeArXeKr arc lamp frames at the beginning or end of each night without moving the instrument configuration.

\begin{figure}
\epsscale{1}
\plotone{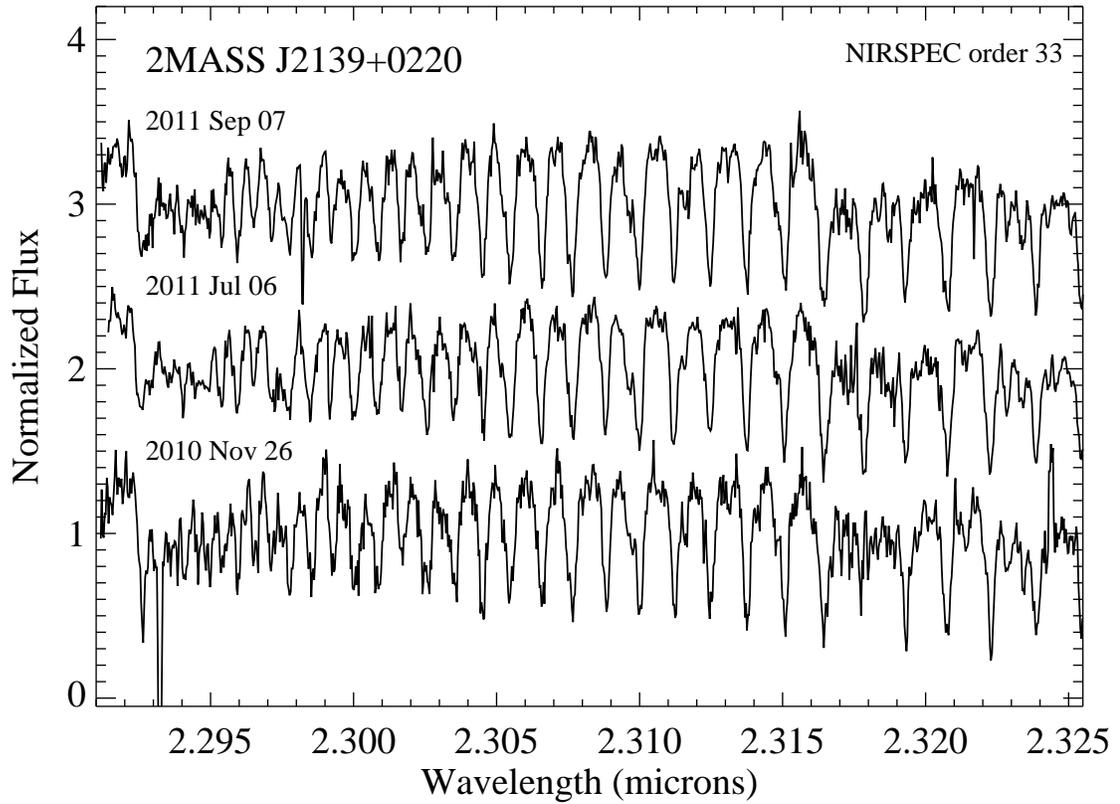}
\caption{NIRSPEC order 33 spectra for J2139+0220 from (top to bottom) 2011 September 7, 2011 July 6 and 2010 November 26 (UT).  Spectra are corrected for barycentric motion, normalized and offset by integer constants.}
\label{fig:nirspec}
\end{figure}

Data were reduced using the REDSPEC package\footnote{See \url{http://www2.keck.hawaii.edu/inst/nirspec/redspec/index.html}.}.
We focused our analysis on order 33 (2.291--2.326~$\micron$), which samples the strong CO band around 2.3~$\micron$ \citep{2008ApJ...678L.125B}.  Image rectification and wavelength calibration (second order polynomial) were performed using the arc lamp images, and for each observation, wavelength solutions were corrected for barycentric motion.  Flux calibration and telluric absorption corrections were determined from the A0~V spectra assuming a 9480~K blackbody.  
Figure~\ref{fig:nirspec} displays the three epochs of NIRSPEC data for J2139+0220, corrected for barycentric motion.  Signal-to-noise (S/N) for these data are roughly 30--40, and show strong molecular absorption arising from $^{12}$CO 2-0 and 3-1 transitions.  
Radial velocities based on these observations are discussed in Section~4.5.

\section{Photometric Monitoring Observations}

\subsection{Aperture Photometry}

\begin{figure}
\epsscale{0.75}
\plotone{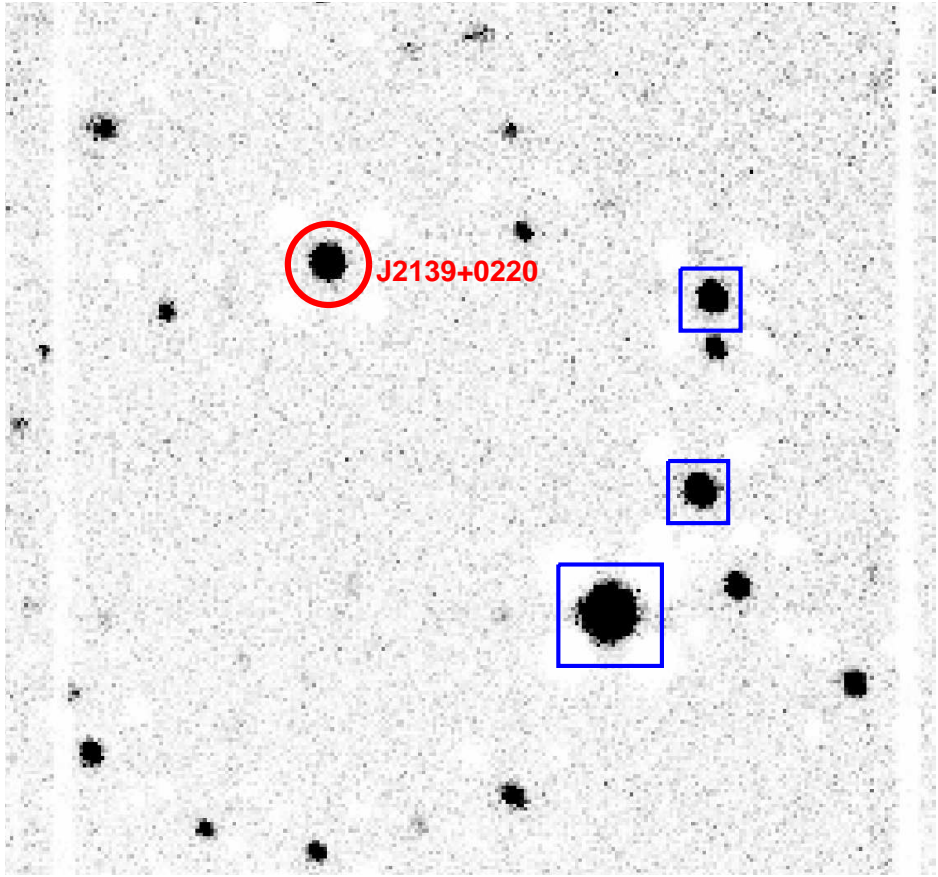}
\caption{Master Gemini image of J2139+0220 for data taken 2010 October 27. The field of view is 2$\farcm$9$\times$2$\farcm$9 with North up and East toward the left.  J2139+0220 is indicated by the red circle, while comparison stars for differential photometry are marked in blue squares. The four ghost images around each source are due to the imperfect median combination of the overall sky frame.}
\label{2139master}
\end{figure}

Aperture photometry was measured for all visible sources in each of the data frames. Centroids were found using the \textit{centroid} routine, a variant of the \textit{find} routine for the DAOPHOT package \citep{1987PASP...99..191S}. Measurements were obtained in two passes. First, sources were located in the first image of a given sequence, and their pixel positions were used as initial guesses for subsequent images based on the dithering sequence, allowing for a drift of up to 5 pixels. In some cases, offsets between images were adjusted manually. These centroids were used to shift, stack and median-combine the full sequence of observations to obtain a high signal-to-noise (S/N) master image. An example master image for 2MASS~J2139+0220 is shown in Figure \ref{2139master}, with the source and comparison star marked. In a second pass, the master image was used to identify dimmer sources, and positions for these were determined by their relative offset from the brighter stars. Aperture photometry was then performed, sampling aperture radii of 2-14 pixels (1.3" - 9.4") and a median background sampled in 10-30 pixel (6.7" - 20.1") annuli. We determined an optimal aperture for a given observation sequence (applied to all measurements of all sources in the sequence) as that which minimized the measurement uncertainty: 
\begin{equation}
\sigma_{flux}^2 = F_{int} + A\left [ (\sigma_{back})^2 + (RN)^2 \right ],
\end{equation}
where $F_{int}$ is the integrated counts (DN) in the aperture, corresponding to shot noise; $A$ is the fractional number of pixels in the aperture, $\sigma_{back}$ is the standard deviation in background counts, and $RN$ = 2.1 DN is the read noise of the detector. We ultimately adopted a 4-pixel aperture, which was found to minimize uncertainties for nearly all of the observations in both datasets. We then averaged the count measurements (equivalent to integrated fluxes) within a dither sequence using uncertainty-weighting.  We also computed the average seeing in each image from the widths of gaussian fits to each source point spread function (PSF).

\subsection{Differential Photometry and Variability}

For each detected source, we first computed normalized fluxes, defined here as the individual flux measurement $i$ divided by the time-averaged mean over a given observing sequence:
\begin{equation}
\bar{F}_i \equiv F_i/ \langle F \rangle,
\end{equation}
where $\langle F \rangle = \frac{1}{N}\sum_{i=1}^N F_i$. 
Differential magnitudes were computed for each science target (UCD) by dividing its normalized flux by the sum of normalized fluxes of all comparison stars (C) in the Gemini FOV:
\begin{equation}
\Delta = -2.5log_{10}\left (\frac{\bar{F}_{UCD}}{\frac{1}{N} \sum_{i}^{N} \bar{F}_{C,i}}  \right ).
\label{diffmag}
\end{equation}
Uncertainties were computed as 
\begin{equation}
\sigma_{\Delta}^2 = 1.179\left [\left ( \frac{\bar{\sigma}_{UCD}}{\bar{F}_{UCD}} \right )^2 + \frac{\sum_{i}^{N}\left (\bar{\sigma}_{F_{ci}}  \right )^2}{\left (\sum_{i}^{N}\bar{F}_{ci}\right )^{2}} \right ],
\label{diffmagerr}
\end{equation}
where $\bar{\sigma} = \sigma/\langle F\rangle$.
\citet{2008PASP..120..860B} utilized a similar equation for computing their differential magnitudes and uncertainties, but without normalized fluxes. The comparison stars were initially chosen as all detected sources in the FOV excluding the science target. 

Typical uncertainties in differential magnitudes ranged from 1\% to of 5\% for most sources. We conservatively introduce a 1\% error in quadrature with these values to account for systematic uncertainties associated with, e.g., flat fielding, OH sky line variations, and other second-order atmospheric opacity effects.

\begin{figure}
\epsscale{1}
\plotone{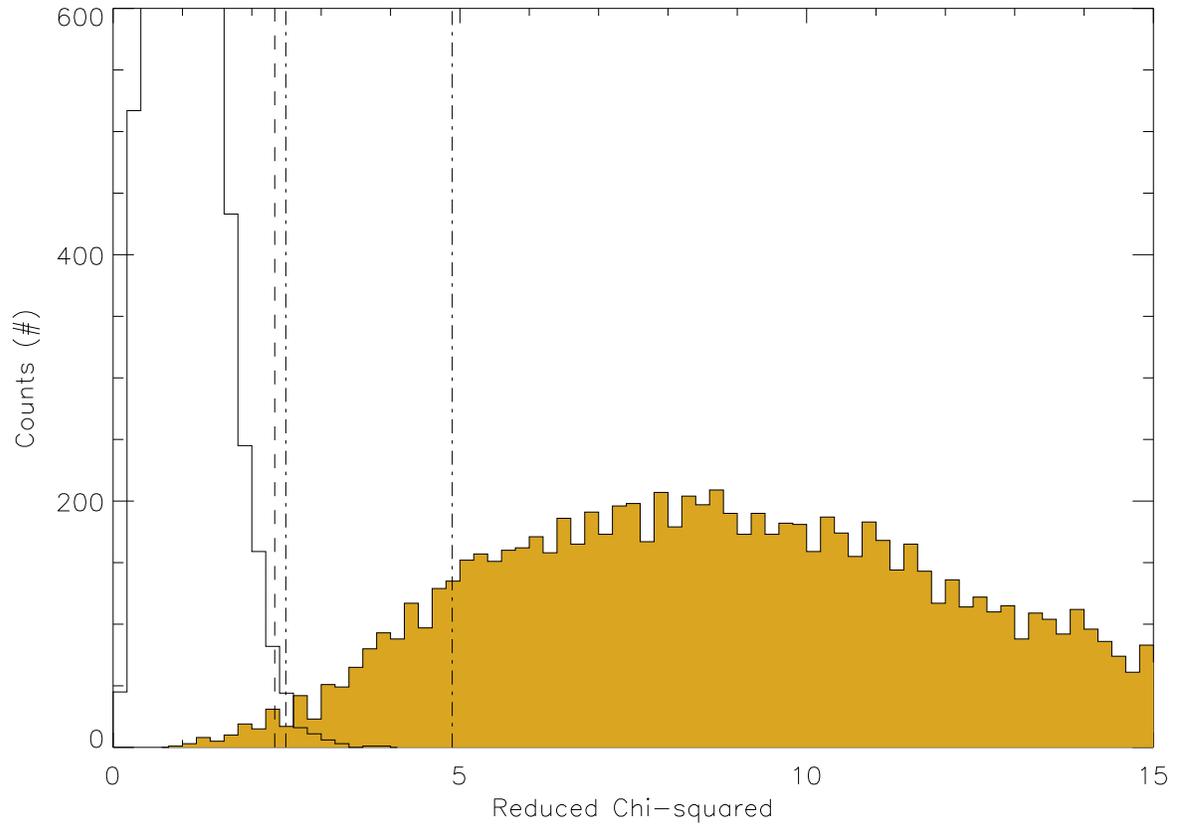}
\caption{Monte-Carlo simulation of variability for a signal-to-noise ratio $S=3$. Two histograms representing a non-variable signal (unshaded) and a variable signal(gold shaded) are over plotted. The dashed line represents the top $1\%$ of the non-variable signal and the dash dot lines represent the bottom $1\%$ and $10\%$ of a variable signal (left and right respectively).}
\label{chi-squared gaussian}
\end{figure}

We then measured the reduced $\chi^2_r$ of each source's light curve over the observing sequence:
\begin{equation}
\chi^2_r = \frac{1}{N-1}\sum_{i=1}^{N}\frac{\left ( \Delta - \left \langle \Delta \right \rangle \right )^2}{\sigma_{\Delta} ^2}.
\label{chisq}
\end{equation}
To distinguish truly variable sources, both among the science targets and comparison stars, we determined a threshold $(\chi^2_r)_{min}$ to exclude false positives, defined here as where the highest 1\% of $\chi^2_r$ values for a non-variable light curve with noise $\sigma$ exceeded the lowest 10\% (or 1\%) of  $\chi^2$ values for a variable light curve with semi-amplitude $S\sigma$.  
Distributions of $\chi^2_r$ values were generated through simulated data via Monte Carlo methods, assuming gaussian noise, 1 $< S <$ 10, and sampling light curves identically as our observations.   
We found that a minimum threshold was reached only for $S \gtrsim$ 3, which defines our minimum detection threshold for each target in units of average uncertainty, and that $(\chi^2_r)_{min} > 1.5$ (2.0) corresponds to a bona-fide variable source at the 90\% (99\%) confidence level.  The Monte-Carlo simulation to demonstrate the thresholds is shown in Figure \ref{chi-squared gaussian}. The 90\% level for the non-variable signal is $\chi^2_r = 1.5$ and the overlapping point with the bottom 1\% of the variable and the top 1\% of the non-variable signals is approximately $\chi^2_r = 2.27$.

In addition to comparing $\chi^2_r$ to $(\chi^2_r)_{min}$ for all science targets, we checked that all comparison stars had $\chi^2_r < 1.5$; any exceeding this threshold were rejected (to a minimum of 2 sources), and differential magnitudes recomputed.  

\begin{figure}
\epsscale{1}
\plottwo{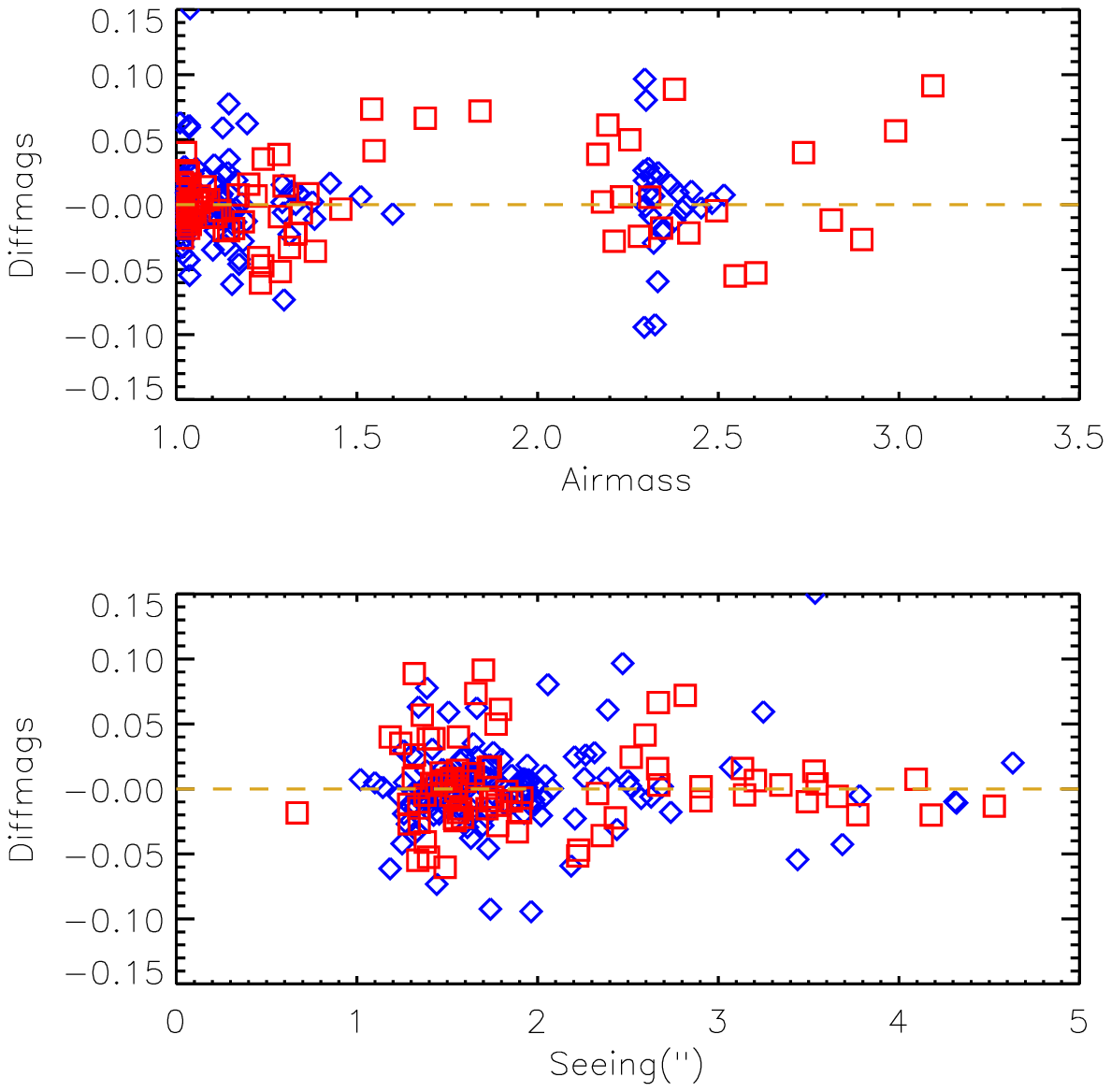}{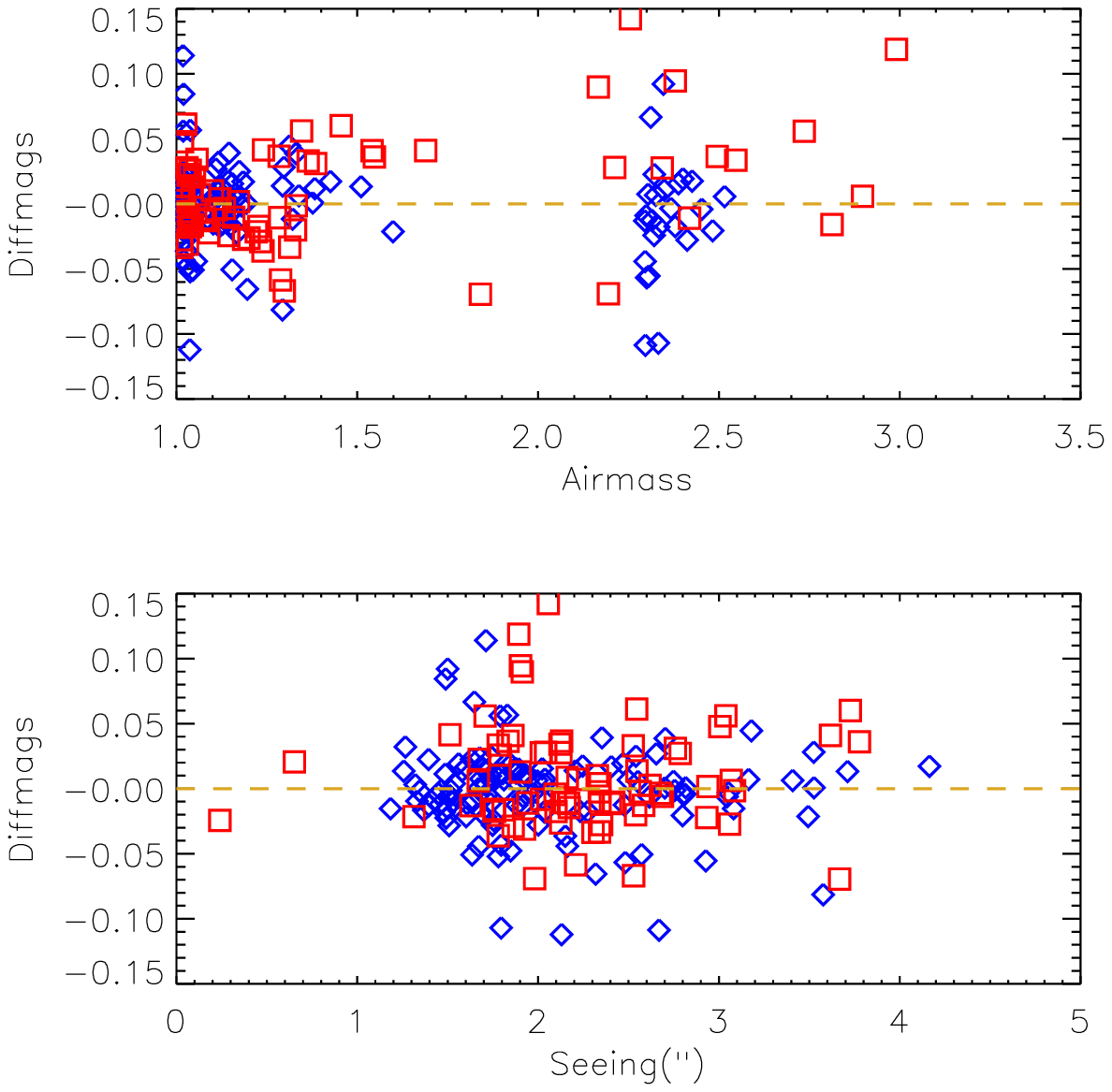}
\caption{Normalized differential magnitudes (differential magnitudes minus mean over an observing sequence) versus airmass (top) and seeing (bottom) for all observations and nights in J-band (left) and K-band (right). L-type dwarfs are plotted as diamonds while T-dwarfs are plotted as squares. We find no significant trends with these parameters for either spectral class.}
\label{varcheck}
\end{figure}

We also checked for trends in differential magnitudes with airmass or seeing variations.  As the spectral energy distributions of our sources are generally distinct from background stars, wavelength-dependent changes in telluric opacity could potentially give rise to false detections of variability in broad-band photometry (e.g., \citealt{2003MNRAS.339..477B}).  Figure~\ref{varcheck} plots differential magnitudes (minus sequence mean) versus estimated seeing and airmass in J- and K-band.  
In the J-band we found correlation coefficients of 0.007$\pm$0.009 versus seeing and 0.003$\pm$0.004 versus airmass for all dwarfs, indicating that any variation in telluric opacity was spectrally grey. For the L-dwarfs, we found correlation coefficients of 0.004$\pm$0.005 (seeing) and 0.008$\pm$0.010 (airmass), and for the T-dwarfs we found coefficients of 0.02$\pm$0.03 (seeing) and 0.005$\pm$0.006 (airmass), again indicating no statistically significant trends. Similar results were found for the $K^{\prime}$ data.

\subsection{Results: Variable Sources}

\begin{figure}
\epsscale{0.75}
\plotone{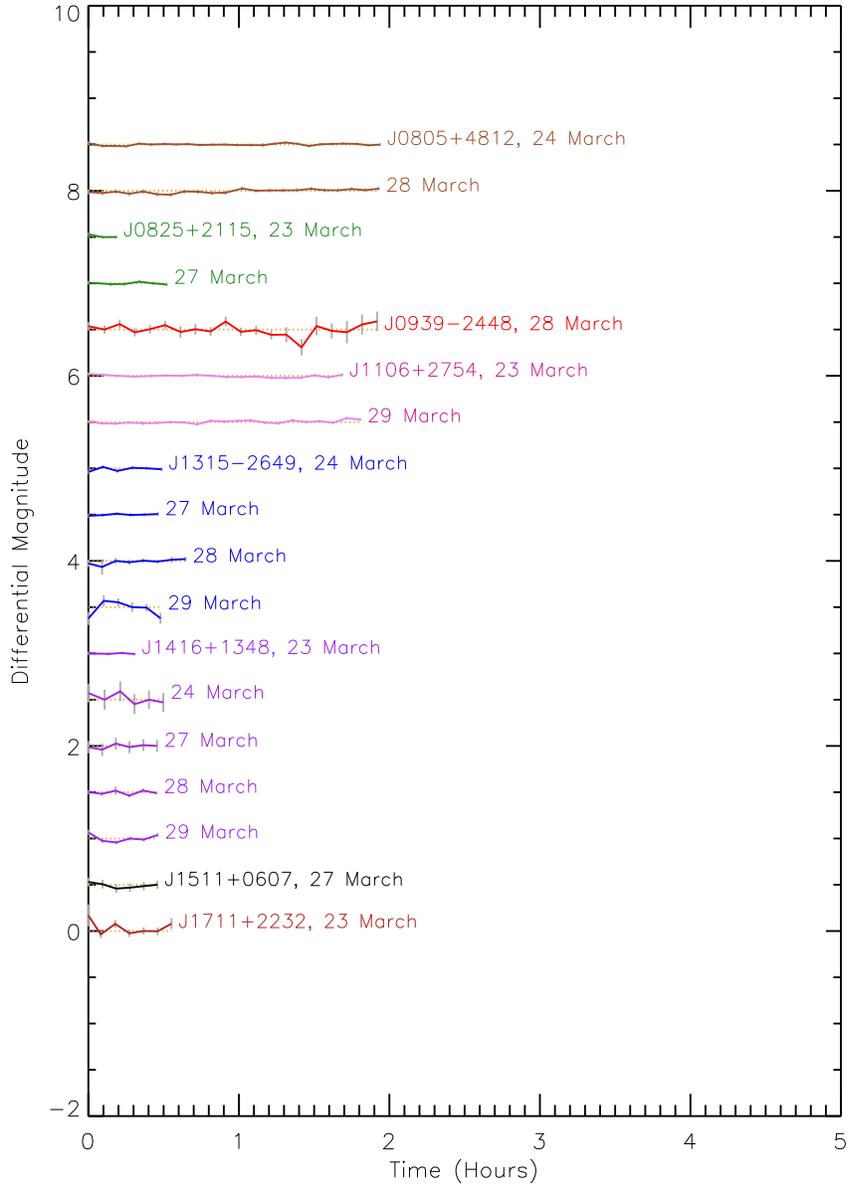}
\caption{$J$-band light curves for all sources observed during the March 2010 run. Source name and observation date are listed to the right of each curve; uncertainties are overplotted on the data points in gray. For sources visited on multiple nights, light curves are displayed in the same color as the first night. Median differential magnitudes of the comparison stars for each source are indicated by the dotted lines}
\label{MarchLC}
\end{figure}

\begin{figure}
\epsscale{0.75}
\plotone{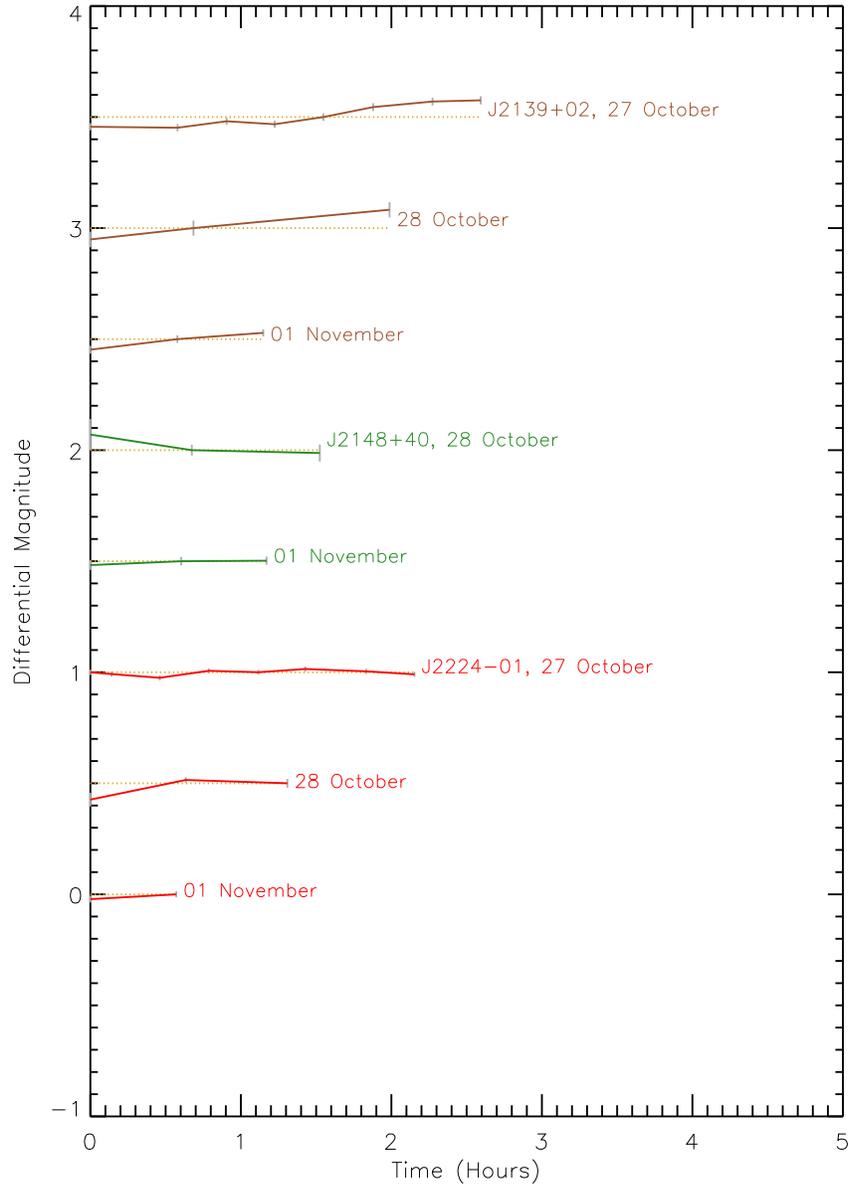}
\caption{Same as figure \ref{MarchLC} for the October 2010 run.}
\label{OctoberLC}
\end{figure}

Table~\ref{resultstable} and table \ref{resultstablek} list the reduced $\chi^2_r$ values, median photometric uncertainties and variability amplitudes of our targets for $J$ and $K$-band data. 
Figures~\ref{MarchLC} through~\ref{OctoberLC} display the differential light curves.
Based on the criteria established in Section~2.4, adopting a 90\% detection threshold ($\chi^2_r > 1.5$), and imposing $\chi^2_r$ of the comparison sources to be less than 1.5, we find that J1315-2649 , J1711+2232, and J2139+0220 show significant variability at $J$ during one or more observing sequences, while J0939-2448 exhibits variability at $K^{\prime}$. 
We also analyzed combined light curves for sources observed over more than one day.  In this case, J2139+0220 was found to be variable over a period of 3 days. Among our variable sources, J2139+0220 and J0939-2448 are the only sources with $S > 3$, with J1711+2232 a marginal case with $S > 2.5$. Each of these sources is examined individually in Section~4.

\subsection{Lomb-Scargle Periodogram Analysis}

For variable sources monitored over periods greater 1 hour, we performed Lomb-Scargle periodogram analyses to search for significant periodicities \citep{1982ApJ...263..835S}. We examined frequencies corresponding to periods ranging from three times the sample rate (i.e., above Nyquist sampling) to one-half the total observation period.  Frequencies were oversampled by a factor of ten. To properly account for uncertainties, we calculated periodograms for 1000 realizations of the data with each differential magnitude adjusted by a normal distribution scaled to the corresponding uncertainty. The final ``raw'' periodogram and its uncertainty was taken as the mean and standard deviation in each frequency bin.

Given the large time gaps in data taken over several nights (e.g., J2139+0220), sampling effects were a concern.  To separate false periods arising from sampling from true periodic variations, a window function was computed by randomly shuffling the observed light curve 1000 times and computing the associated periodograms.  For each frequency bin we determined the maximum value encompassing to 95\% of the amplitude distribution. Dividing the raw periodogram by this window function yields a significance spectrum.
Based on experiments with simulated light curves sampled in the same manner as the data, we determined that peaks in the significance spectrum $>$ 1 correspond to real variations.  These peaks were fit to Gaussian function, whose peak and half-width were used estimates of the variability frequency and its uncertainty.

\subsection{Phase Dispersion Minimization Analysis}

In addition to periodogram analysis, we performed phase dispersion minimization (PDM) analyses on our variable sources using the formalism of \citet{1978ApJ...224..953S}. This technique compares the overall variance ($\sigma^2$) of the light curve to the piece-wise variance of the same curve folded over a given period ($s^2$).   Significant periods appear as a minimum in the ratio $\Theta \equiv s^2/\sigma^2$.
We sampled periods over the same range as the periodogram analysis in steps of 0.01 hours, and light curves were sampled in steps of 0.5 in phase.  We incorporated uncertainties in the measurements in a similar manner as the periodogram analysis, and also examined sampling and aliasing effects by injecting periodic signals corresponding to any minima identified.

\section{Analysis of Individual Variable Sources}

\subsection{J0939-2448}

J0939-2448 is an unusually blue T8 dwarf, exhibiting a K-suppressed spectrum that likely arises from enhanced H$_2$ absorption in a high-surface gravity atmosphere \citep{2005AJ....130.2326T,2006ApJ...639.1095B}. \citet{2008ApJ...689L..53B} and \citet{2009ApJ...695.1517L} noted a mismatch between the near-infrared and mid-infrared spectrum of J0939-2448, and evidence of the source being overluminous, leading both studies to suspect that this source is an unresolved brown dwarf binary. 
Our observations of J0939-2448 over 2 hours on 2010 March 28 (UT) reveal significant variability at $K^{\prime}$ with a semi-amplitude of  0.31 mags ($S$ = 3.3), but no significant variation at $J$. The $K^{\prime}$ variability was the largest  detected in our sample, but is tempered by the relatively low S/N ($\approx$ 4) of the photometry over all dither positions for the entire exposure.  Hence, we characterize this source as a marginal variable requiring confirmation.

\subsection{J1416+1348A}

J1416+1348A is an L6 dwarf \citep{2010ApJ...710...45B, 2010AJ....139.1045S} with a T7.5 spectral type binary companion at a projected separation of approximately 75 AU \citep{2010AJ....139.2448B, 2010A&A...510L...8S, 2010MNRAS.404.1952B}. J1416+1348A is unusually blue in its NIR colors, possibly due to high surface gravity, low condensate opacities, unusually thin clouds with large grain size, and/or subsolar metallicities \citep{2011arXiv1108.4678C, 2012arXiv1201.2465D}. Our observations of this source for 5 nights revealed signs of J-band variability only on the final night, with a semi-amplitude of 0.05 mags, a $\chi^2$ of 2.67 but with an S of 2.35, making this a marginal case. 

\subsection{J1711+2232}

J1711+2232 is an optically-classified L6.5 dwarf \citep{2000AJ....120..447K}, for which near-infrared spectroscopy has revealed signatures of CH$_4$ absorption \citep{2004ApJ...607..499N,2010ApJ...710.1142B}.  This unusual absorption was interpreted in the latter study as an indication that this source is a spectral binary with L5 plus T5.5 components (it is unresolved in HST observations; \citealt{2003AJ....125.3302G}).  Our short sequence of observations (0.7~hr) on 2010 March 23 (UT) reveal signficant variability at $J$-band with a semi-amplitude of 0.10~mags ($S$ = 2.5).  The source also appears quite variable at $K^{\prime}$; however, the comparison field sources are insufficiently stable to confirm this. 
Given the short timescale of the observation, we did not search for periodicity. 

\subsection{J2139+0220}

J2139+0220 is a near-infrared classified T1.5 dwarf with unusually red near-infrared colors for its spectral type. The spectrum of J2139+0220 reveals significant $CH_4$ absorption at 1.4 \micron, with diminished absorption in the H and K-bands \citet{2010ApJ...710.1142B}.

Our observations over three nights in October/November 2010 reveal persistent variability at $J$-band, with semi-amplitudes of 0.062~mag, 0.067~mag and 0.038~mag ($S$ = 3.6, 2.0 and 2.2, respectively).  Combining data from all three nights, we find a overall semi-amplitude of 0.067~mag ($S$ = 3.9).  This was the most significant variation in our sample.

\begin{figure}
\epsscale{1}
\plotone{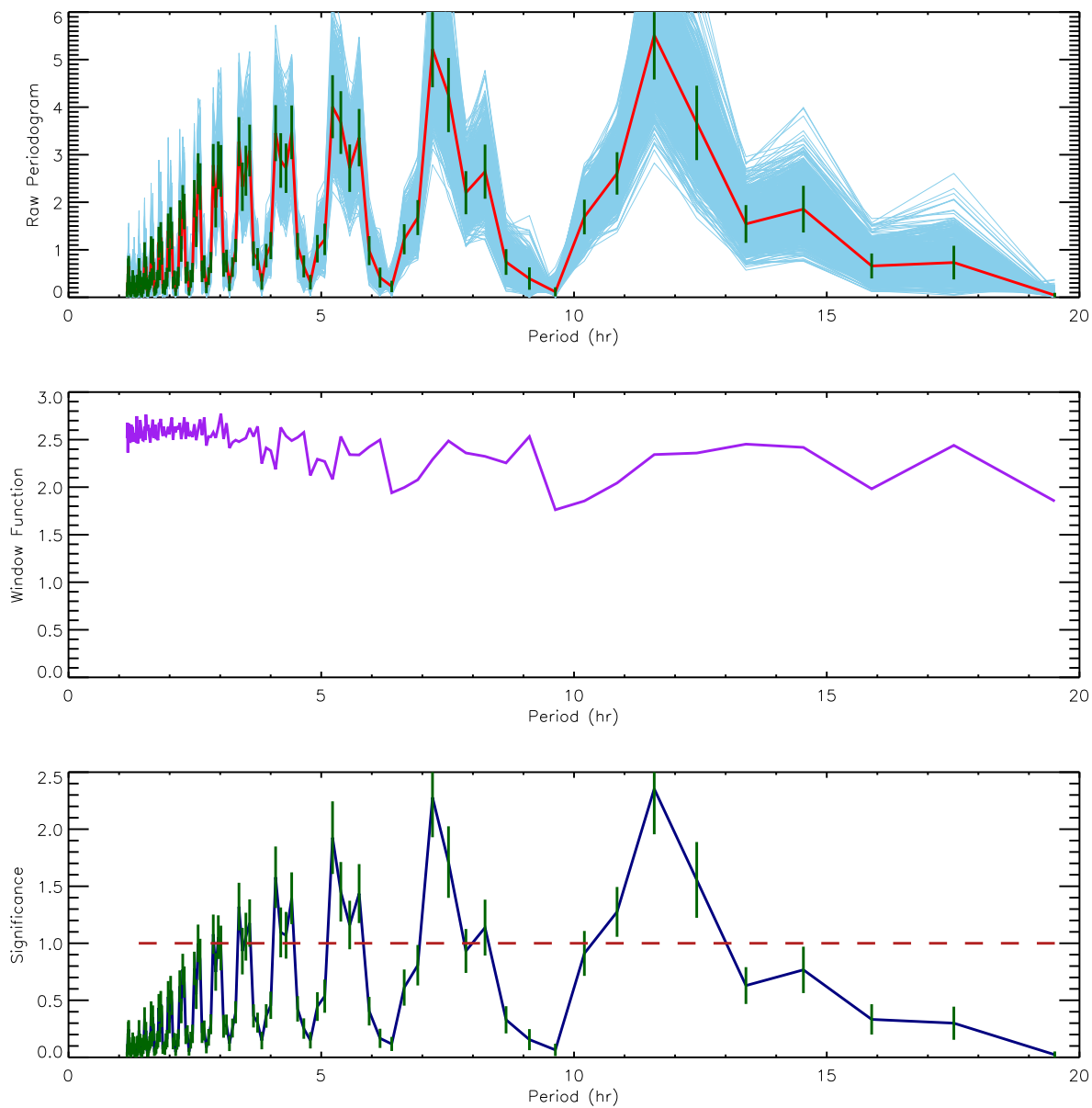}
\caption{J-band Lomb-scargle periodogram analysis of J2139+0220 spanning 3 days. The top panel shows the mean (red line) and uncertainty (green bars) of the raw periodogram based on the Monte Carlo simulation (all included periodograms are overplotted in light blue). The middle panel shows the window function for the 95\% level. The bottom panel shows the significance of the mean periodogram, a significance of 1 is marked by the red dashed line.}
\label{periodogram}
\end{figure}

\begin{figure}
\epsscale{0.9}
\plotone{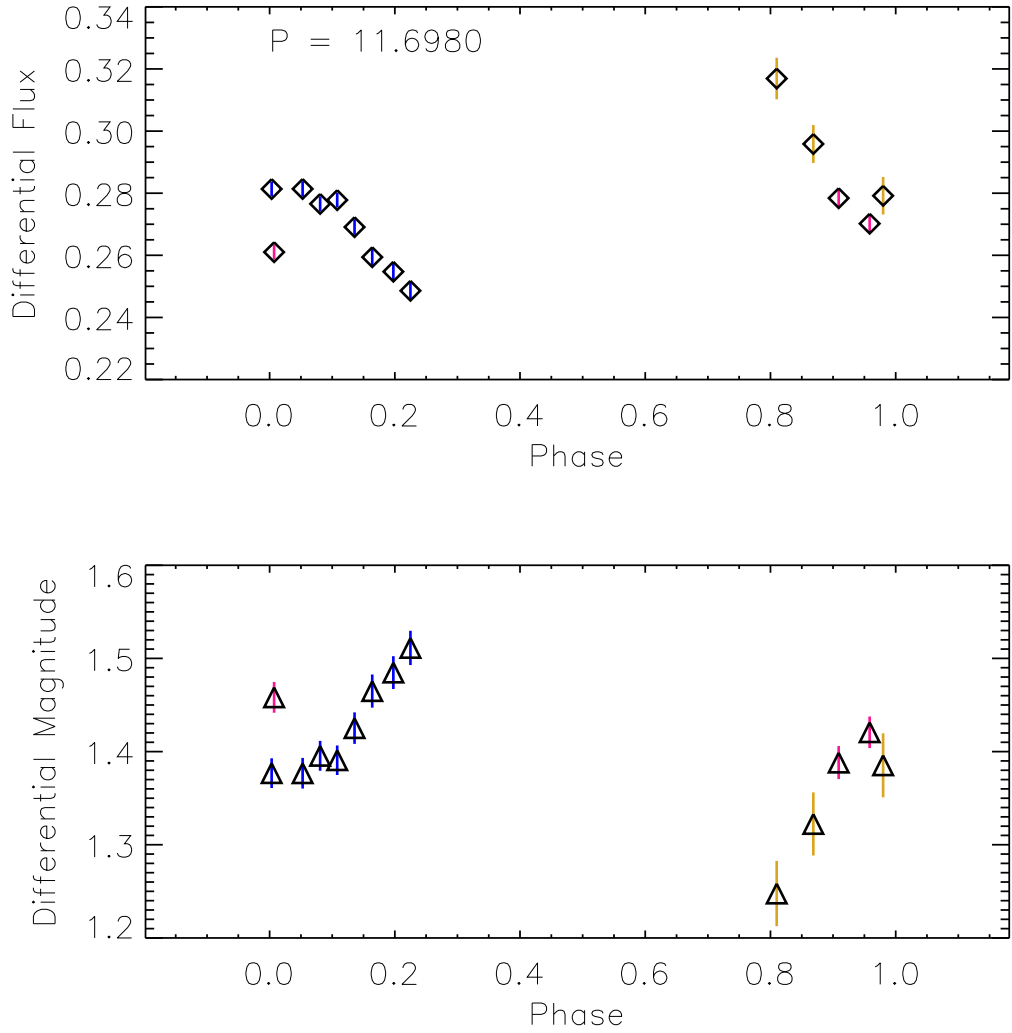}
\caption{Folded J-band light curve of J2139+0220 for all 3 days. The light curve from the 27th of October is shown in blue, the 28th of October shown in gold, and the 1st of November shown in pink. The top figure shows differential flux over time, while the bottom figure shows the differential magnitude over time. The light curve was folded over 11.7 hours.}
\label{ffold11J}
\end{figure}

\begin{figure}
\epsscale{0.9}
\plotone{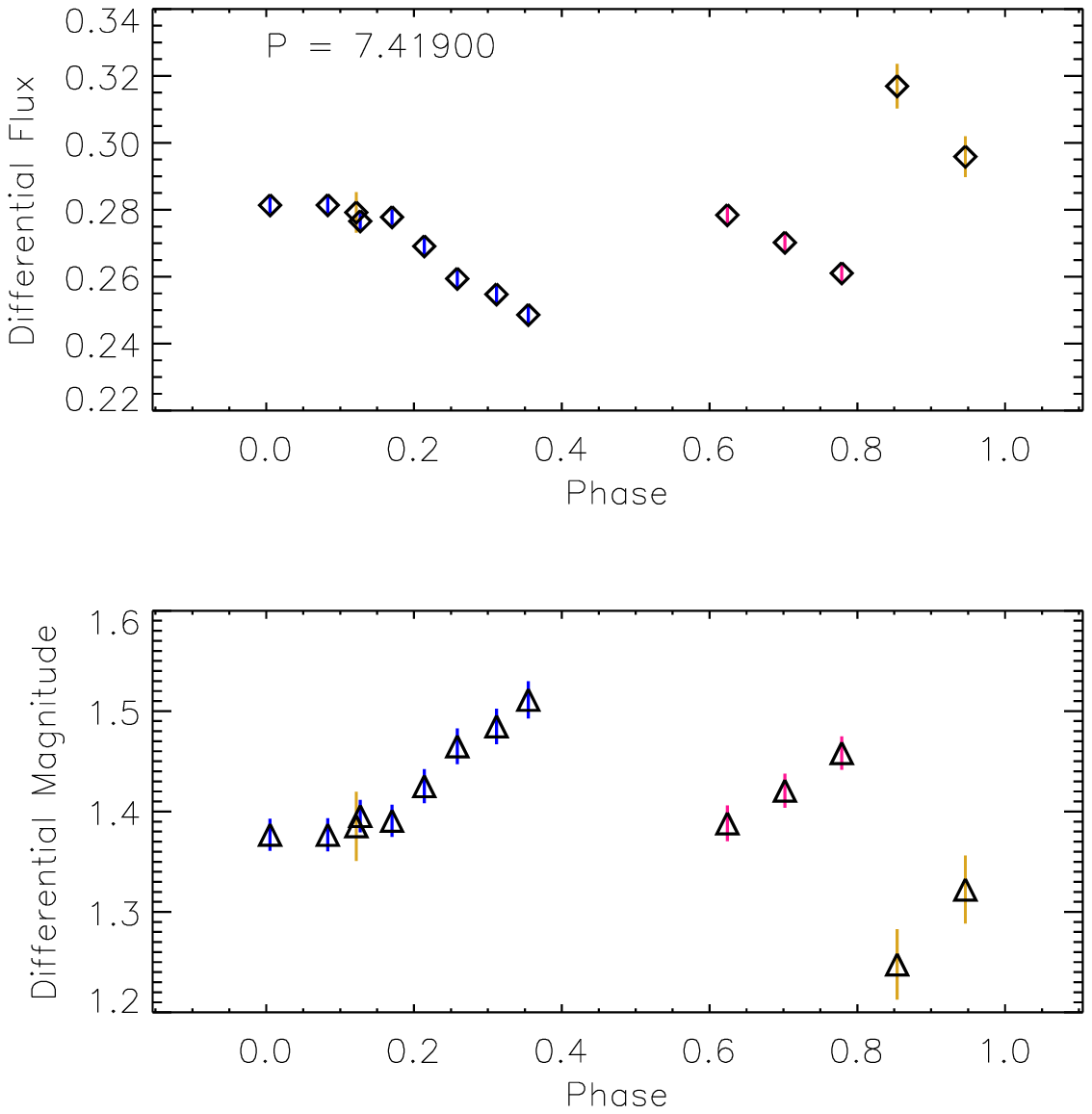}
\caption{Same as figure \ref{ffold11J}.}
\label{ffold7J}
\end{figure}

\begin{figure}
\epsscale{0.9}
\plotone{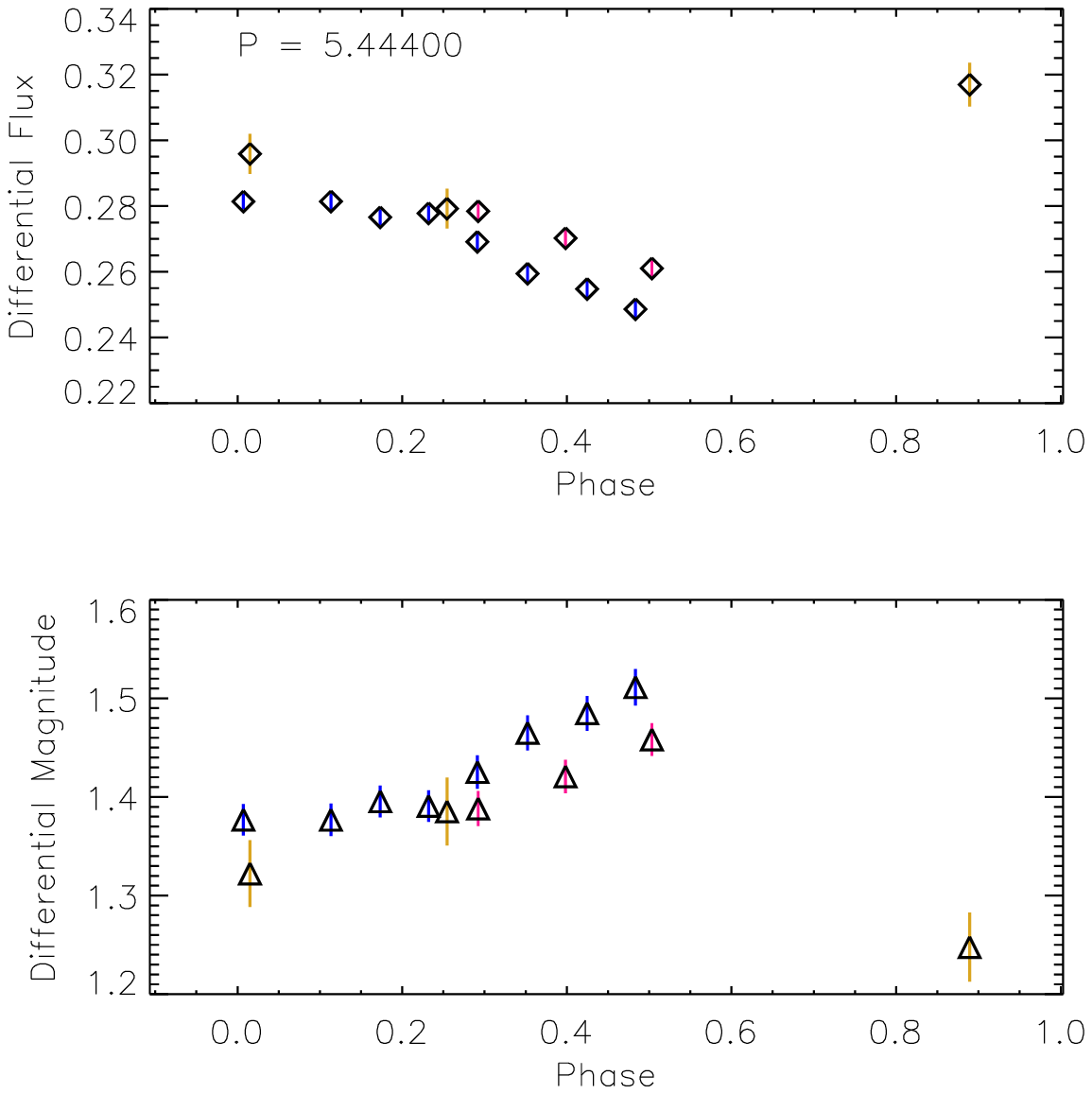}
\caption{Same as figure \ref{ffold11J}.}
\label{ffold5J}
\end{figure}

\begin{figure}
\plotone{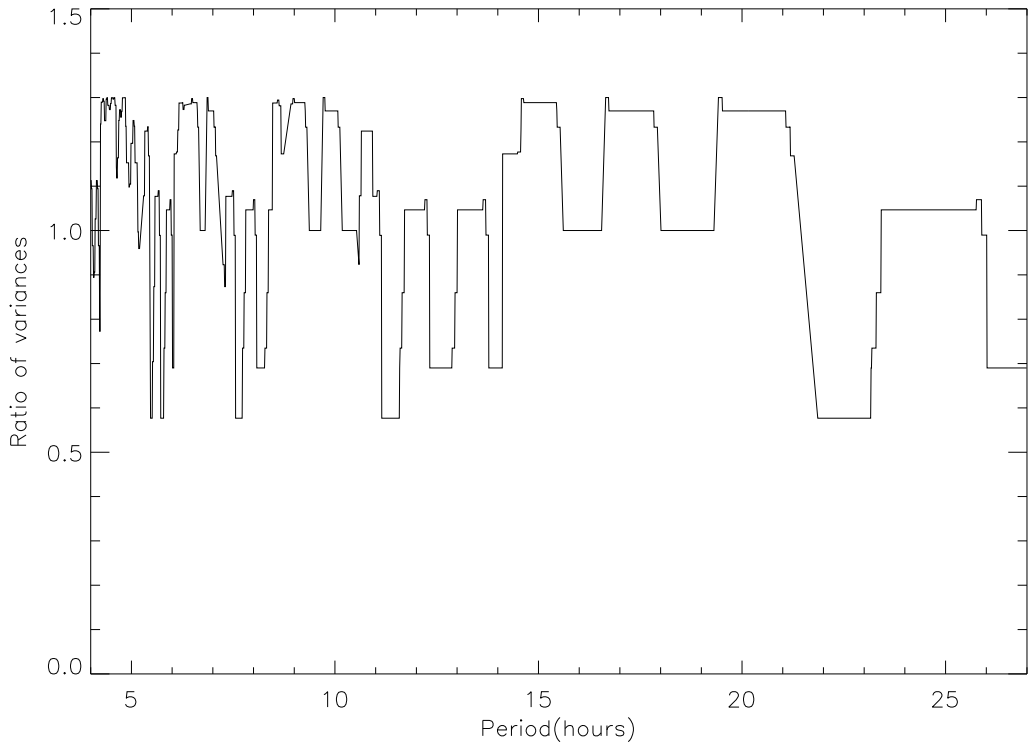}
\caption{Phase Dispersion Minimization results for J2139+0220. Periods in hours are plotted on the x-axis and $\Theta$, the ratio of variances in plotted on the y-axis. Minimum values represent true periods.}
\label{pdm}
\end{figure}

Given its significant variation over a relatively long baseline, we performed both periodogram and PDM analyses on our multi-day light curve.  Figure~\ref{periodogram} displays the results of the former, for which we find significant peaks (1$\sigma$ above significance of 1) at periods of 11.7$\pm$0.5~hr, 7.42$\pm$0.22~hr, 5.44$\pm$0.17~hr and 4.26$\pm$0.12~hr. 
Figure \ref{ffold11J} shows the light curve folded over 11.7~hr, driven largely by the overlap in data from October 28 and November 1 toward the end of the phase cycle.
Figure \ref{ffold7J} shows the light curve folded over 7.42~hr, driven by the overlap of one data point on November 1 with data on October 27.   
Figure \ref{ffold5J} shows the light curve folded over  5.44~hr, which has the most overlap in the night-to-night observations.  
Relative to the this last period, the others are found to be close to 2:1, 4:3 and 3:4 harmonics.
PDM analysis (Figure~\ref{pdm}) also shows similar significant periods at 22.51$\pm$0.054~hr, 11.36$\pm$0.10~hr, 7.64$\pm$0.06~hr and 5.65$\pm$0.05~hr, which have ratios near 4:1, 2:1 and 4:3 about the shortest of these periods.  

The structure of the periodogram peaks reveals harmonics from peak-to-peak but also reveals double peaks for each significant point. Similarly in the PDM, the troughs are tripled for each significant point. In both cases, this is due to the sampling of the data, as the peaks themselves are not harmonics of each other. The separation between the periodogram peaks is approximately 0.3 hours for the shorter period peaks, to approximately 2.5 hours for the longer period peaks. The difference in the troughs of the PDM for each triple is between 0.5 hours for the shorter periods, to 1 hour for the longer periods, which indicate sampling issues as observations were spaced out every 30 minutes.

\begin{figure}
\plotone{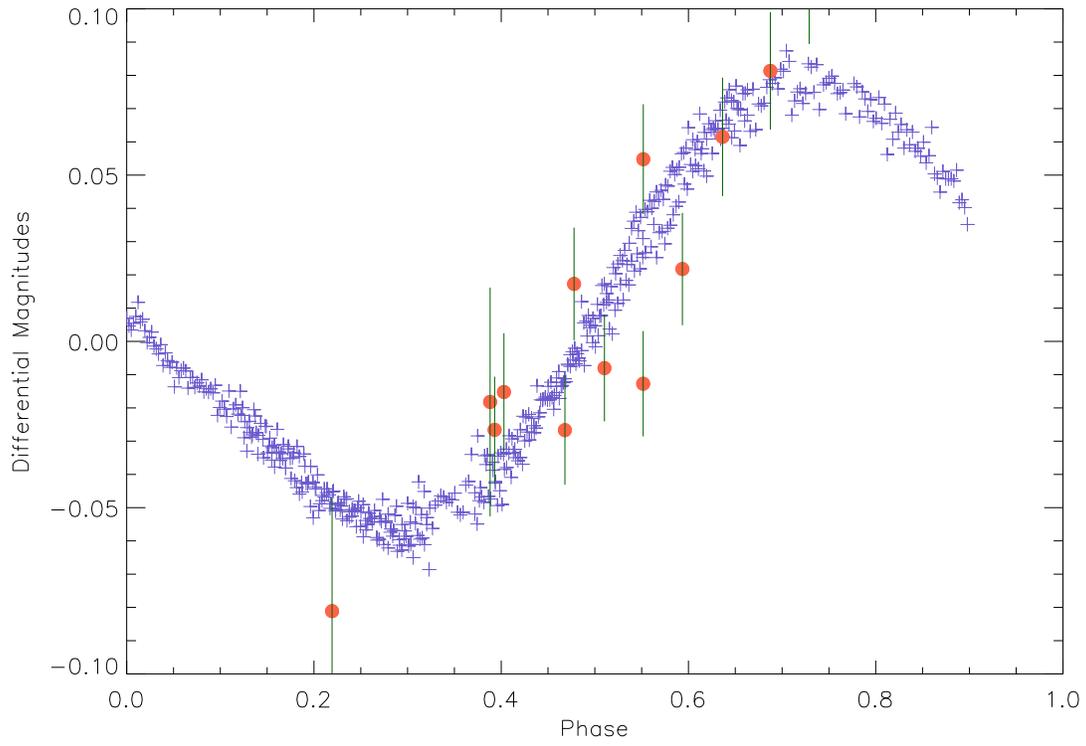}
\caption{Folded flux light curve of J2139+0220 with data from Radigan et. al overplotted in blue with the $+$ symbol, and our data in red circles phased to 7.72 hours, with a phase offset $\phi$ of 0.388, and a flux scaling ratio of 0.46 applied.}
\label{2139jackie7.72ffold}
\end{figure}

J2139+0220 has been independently identified as a relatively large-amplitude variable by \citet{2012ApJ...750..105R} based on several days of continuous near-infrared monitoring.  That study finds a semi-amplitude of 26\% at J-band---the largest detected to date---with a precise periodicity of 7.721$\pm$0.005 hours.  This period is statistically consistent with the 7.42~hr and 7.64~hr periods found in our data, and Figure~\ref{2139jackie7.72ffold} compares our data with that from \citet{2012ApJ...750..105R} phased to 7.721~hr. We found a phase offset between our data and the Radigan et al. study of $\phi = 0.39$. 

The Radigan et al.\  study also finds their light-curve to vary both in amplitude night-to-night and phase over several periods, which they interpret as arising from evolving cloud features. The amplitude of our variations of 0.067 mag is considerably smaller than the 0.15 mag amplitude reported in the Radigan study. To assess if this difference is true variability or instrumentation and filter effects, we calculated the difference in relative brightness of this sources between the two filters used. We modeled the Gemini J-band filter as a box function with boundaries of 1.11 to 1.39 \micron.  According to the \citet{2012ApJ...750..105R} study, the WIRC J-band filter was similar to the Mauna Kea Observatory (MKO) J-band filters, which we used here. These filters were convolved with the best-fit combination of cloudy and clear atmospheres from \citet{2008ApJ...689.1327S} that Radigan et. al found to reproduce the variability amplitudes at different wavelengths. The amplitudes were found to be the same for both filters indicating that the difference in the observed amplitudes is possibly due to intrinsic variability and changes in the cloud structure.

\begin{figure}
\epsscale{0.9}
\plotone{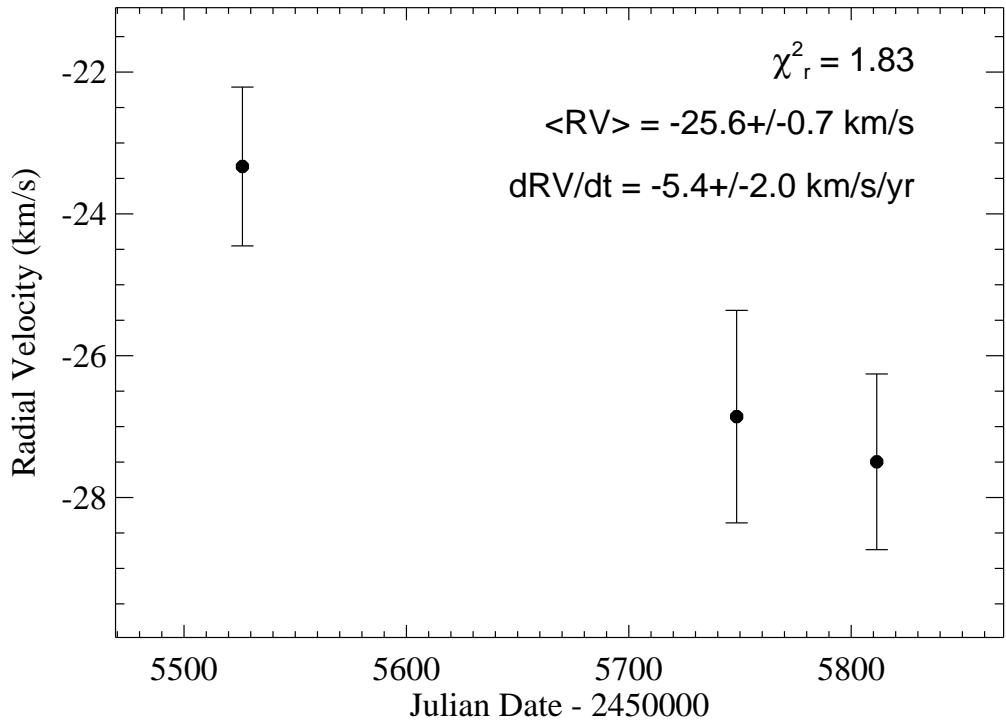}
\caption{Radial velocity of J2139+0220 versus Modified Julian Date, spanning November 2010 to September 2011. The reduced $\chi^2$ of the velocities, as well as the average radial velocity and radial velocity trend slope are listed. A constant -25.6 km~s$^{-1}$ radial velocity is the best fit case.}
\label{fig:rv}
\end{figure}

Interpretation of these data must take into account the possibility that J2139+0220 is a tight binary system since irradiation effects, magnetic interaction, or other orbital interactions could drive surface asymmetries.  We tested this hypothesis with our NIRSPEC observations.  We cross-correlated each of the three NIRSPEC spectra with equivalent data for the T2 dwarf and radial velocity standard SDSSp J125453.90-012247.4 \citep{2000ApJ...536L..35L,2007ApJ...666.1205Z}
following the methodology of \citet{2012AAS...22032810B}.  The resulting velocities are listed in Table~\ref{tab:rv} and plotted in Figure~\ref{fig:rv}.
These data are marginally consistent with a constant velocity of $-$25.6$\pm$0.7~{\kms} ($\chi^2_r = 1.83$). A slight trend of $-$5.4$\pm$2.0~{\kms}~yr$^{-1}$ also fits the data, but this is not statistically significant.  Small variations in the observed radial velocity of this source could arise from the same surface features causing the light variations. Assuming a single spot covering 26\% of the surface as estimated by \citet{2012ApJ...750..105R}, radial velocity variations of up to 3~{\kms} could be discerned if $v\sin{i} \approx$ 10~{\kms} \citep{1997ApJ...485..319S}.  In any case, we can definitively rule out the kind of short-period binary system that would give rise to interaction effects.
Assuming that this interaction is magnetic in origin, so that the system is composed of two $m$ = 0.05~\msun brown dwarfs separated by at most twice their magnetosphere radii ($a$ $\approx$60~R$_{Jup}$ = 4.3$\times$10$^6$~km $\approx$ 0.03~AU; \citet{2009ApJ...699L.148S}) we can determine the maximum semi-amplitude of radial velocity via:

\begin{equation}
K_1 = \left ( \frac{\pi G m}{2P} \right )^{\frac{1}{3}}\frac{sin(i)}{\sqrt{1-e^2}}= \left ( \frac{G m}{2a(1-e^2)} \right )^{\frac{1}{2}} sin(i)
\end{equation}

where we assume the eccentricity $e$ = 0 and the inclination $i$ = 90$\degr$; this yields , we obtain $K _1 \approx 27$ \kms. Assuming a 4$\sigma$ limit on the observed radial velocity variability amplitude of $\approx$ 3~\kms, this impies a maximum inclination $i_{max}$ = 6.2$\degr$ if the variations are caused by binary interactions. The geometric probability of alignment for such a small separation $P(i) = 1-\cos{i}$ = 0.0058 is sufficiently small to be deemed improbable. Indeed the 3~\kms variability limit argues against a near-equal mass binary out to separation of 0.5~AU at 90\% confidence limit. The period for this outer separation limit is 6~days; if we consider a binary whose orbit is synched to a period of 7.6 hours ($\approx 0.0042$~AU or $\approx 8.83~R_{Jup}$), $K_1 \approx 72.5$~\kms, implying a maximum inclination of 2.4$\degr$, and we determine a probability of alignment $P(i) = 1-\cos{i} = 0.00085$.

The apparently single nature of J2139+0220 does not necessarily contradict its identification as a spectral binary candidate, however, as the dual spectral appearance may arise from distinct cloudy and non-cloudy regions on its photosphere. It is possible that other spectral binaries identified in \citet{2010ApJ...710.1142B} may turn out to be similarly variable single L/T transition dwarfs with patchy atmospheres; 2MASS 1711+2232 may be an example of such a source.

\section{Discussion}  

\begin{figure}
\plotone{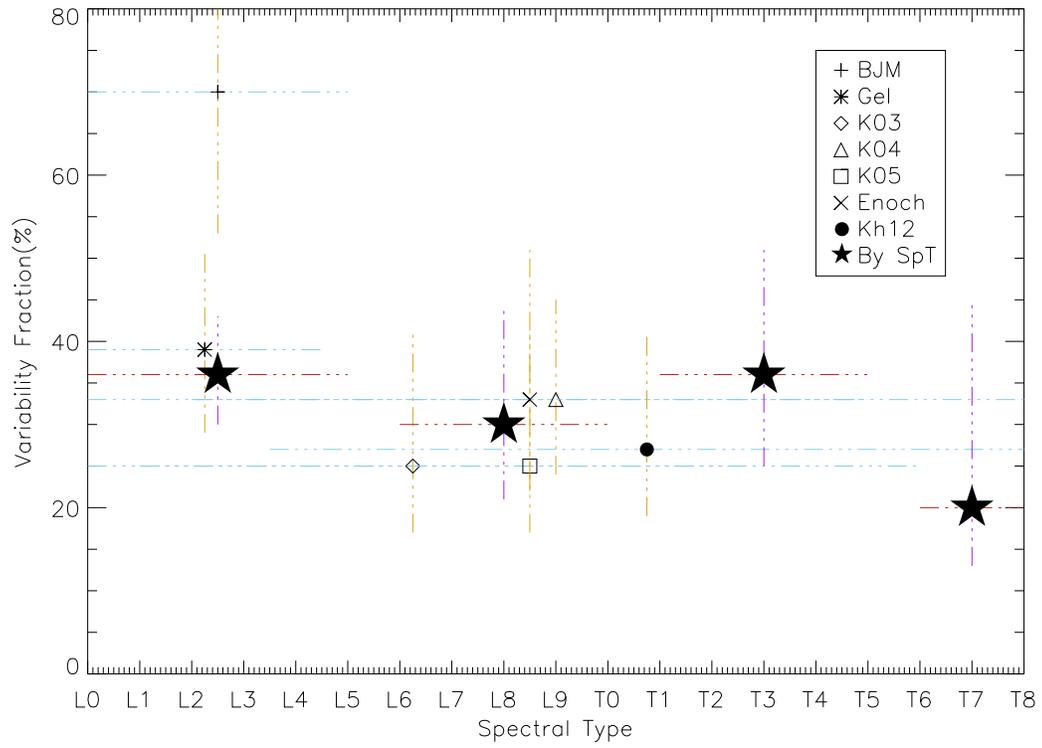}
\caption{Variability rates over spectral type range for the studies listed in Table \ref{tab:var2}. We also indicate the mean and standard deviation for each spectral type range as stars.}
\label{vartableplot}
\end{figure}

We have identified 4 variable L and T dwarfs in a sample of 15. This implies a rate of variability of $27^{+14}_{-8}\%$ with 1$\sigma$ errors based on binomial statistics. Table \ref{tab:var2} and Figure~\ref{vartableplot} show how this variability rate compares to other ``large'' samples of 9 targets or more.  Although the studies were conducted with different filters and report different sensitivity limits, with the exception of \citet{2001A&A...367..218B} all agree with a variability fraction of 25--40\%.  For the 78 brown dwarfs in these 7 studies, 27 sources were variable, giving an overall variability fraction of $35{\pm}5\%$.   The $\gtrsim$70\% rates reported by \citet{2001A&A...367..218B} and \citet{2006A&A...448.1111R} from a sample of 10 L dwarfs are considerably higher and in contradiction with the results of \citet{2002ApJ...577..433G}, who observed a larger sample to the same depth and with a similar filter but found a variability rate of only  39$^{+12}_{-10}$\%.  In addition, we have examined the variability rate for distinct spectral ranges (Table \ref{tab:var2}), and find that the variability rate for each subset agrees with the overall rate of variability. We therefore conclude that the high variability fraction is likely spurious, and that existing data show no clear evidence of a change in the variability rate as a function of spectral type, even for the L/T transition.  Nevertheless, it remains true that the largest amplitude variables detected so far---SIMP J013656.5+093347 \citep{2009ApJ...701.1534A} and J2139+0220---are early-T dwarfs which show evidence of significant cloud structure.

Improving on current statistics is challenging as current samples of known brown dwarfs are mostly located at high Galactic latitudes, where the number of field stars diminishes, making differential photometric analysis more difficult. As seen in Table \ref{resultstable}, the number of comparison stars we were able to use was typically less than 5 and in some cases only 1 or 2. Coupled with the inability to detect small levels of variability (0.01 magnitudes and below), it is clear that infrared variability studies could be improved with larger field-of-view instruments. An instrument with an increase in linear size of 3 ($\ge9'$) would in principle increase the number of reference stars 10-fold. To our knowledge, there are no such instruments available on small ($<$~3m) telescopes. Instruments such as Flamingos, UKIRT \citep{1980Endvr...4..132H}, NEWFIRM \citep{2003SPIE.4841..525A}, and IRIS 2 \citep{2000SPIE.4008.1084G} on smaller telescopes would allow for more dedicated surveys of variable brown dwarfs.

\section{Summary}
We have identified 4 variable L and T dwarfs in a sample of 15 observed: J1416+1348A , J1711+2232, and J2139+0220 in the J-band;  and J0939-2448 in the K-band. Two of the variable sources were spectral binary candidates, and our detection of J2139+0220 confirms both the variability and period of this source as recently reported by Radigan et al.\ (2012), although we find that its variability amplitude has declined since its initial detection.  We also find that this source has a stable radial velocity, suggesting that its spectral binary nature may derive from a patchy atmosphere rather than two distinct sources. We find a variability rate of $27^{+14}_{-8}\%$ for our study and after examining 7 other similar studies (including our own), we report an overall percentage of variability of $35\pm 5\%$ for L and T type brown dwarfs. The challenge in observing low levels of variability necessitates the use of large field-of-view instruments on smaller and more dedicated telescopes.

\section{Acknowledgments}
The authors acknowledge helpful discussion with Kim Griest and Marc Rafelski, and thank Lick Observatory staff Keith Baker, Wayne Earthman, Elinor Gates, Bryant Grigsby and Donnie Redel for their support with the observations. The authors thank Jacqueline Radigan for generously sharing her spectral models with us for the J2139+0220 analysis. We thank the reviewer for constructive comments which helped to strengthen the manuscript. 
AJB acknowledges funding support by the Chris and Warren Hellman Fund.
This work was based on observations made with the 3m telescope at the Lick Observatory in Mt. Hamilton, CA. 

\clearpage

\clearpage

\begin{figure}
\epsscale{0.75}
\plotone{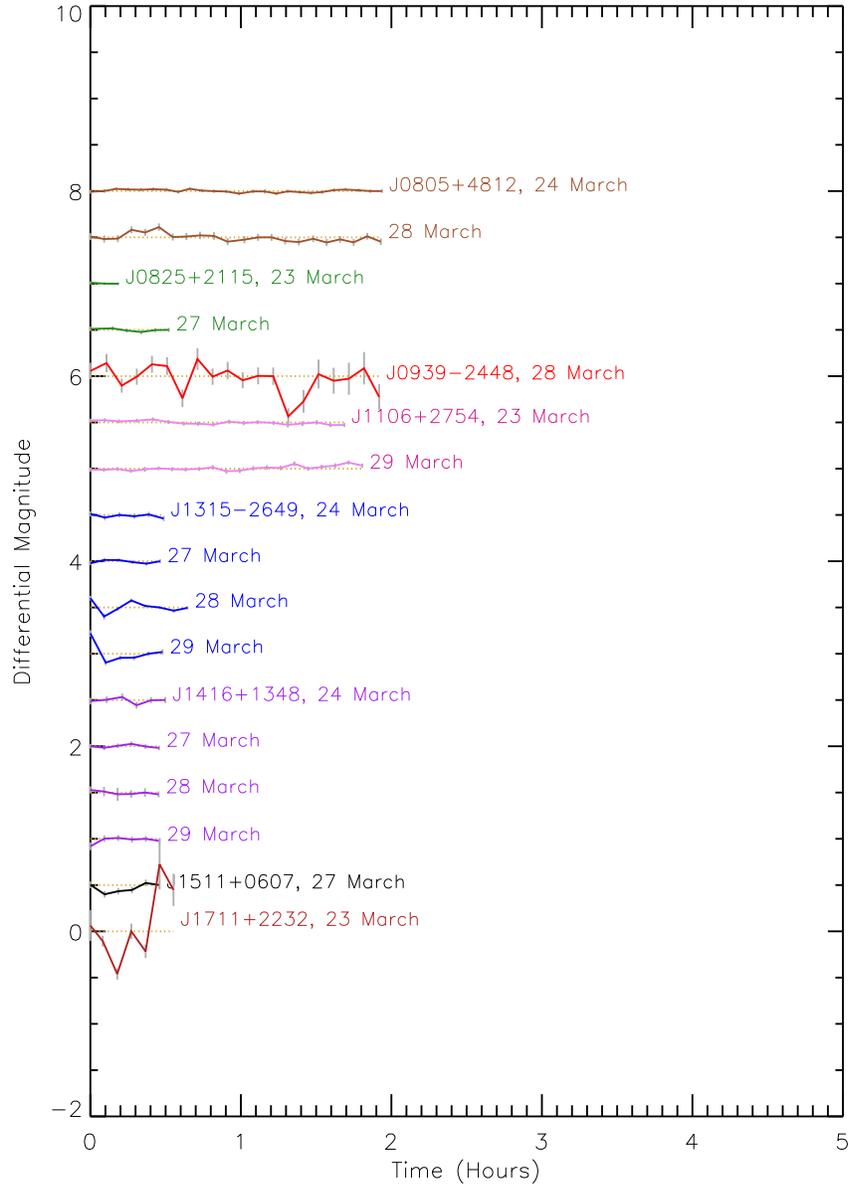}
\caption{Same as figure \ref{MarchLC} but $K$-band light curves for the March 2010 run.}
\label{MarchLC-K}
\end{figure}

\begin{figure}
\plotone{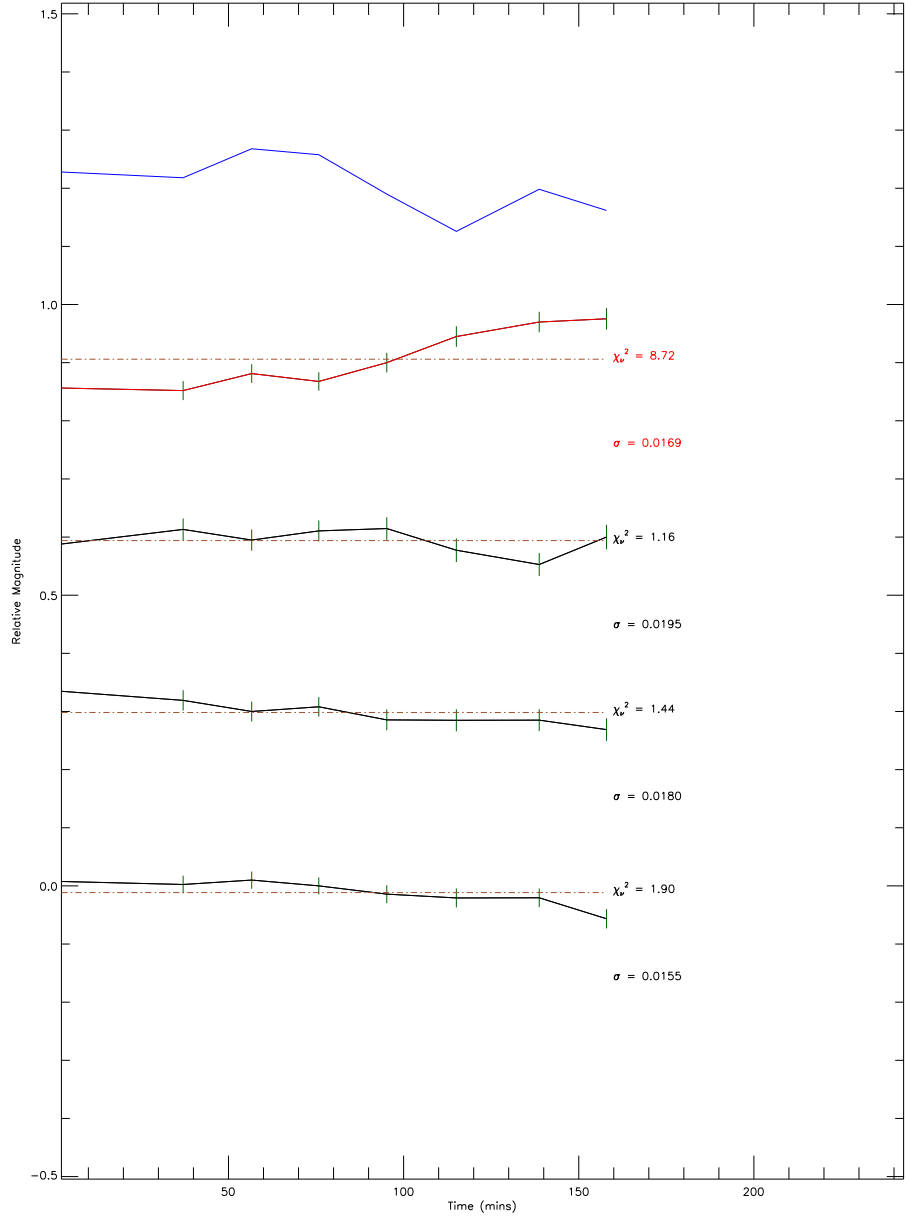}
\caption{Light curve of J2139+0220. The blue line at the top represents the total flux of the comparison stars, and is used to understand the general change in the atmospheric transparency throughout the course of the night. The uncertainties and the reduced chi-squared are listed.}
\label{2139LC}
\end{figure}

\clearpage

\begin{deluxetable}{llllcccll}
\tablecaption{Observational Sample \label{sourcetable}}
\tabletypesize{\scriptsize}
\tablewidth{0pt}
\tablehead{
\colhead{Source} & 
\colhead{Coordinates} & 
\colhead{Optical} & 
\colhead{NIR}& 
\multicolumn{3}{c}{2MASS} & 
\colhead{Type\tablenotemark{a}} & 
\colhead{Refer.}   \\
& & \colhead{SpT} & 
\colhead{SpT} & 
\colhead{$J$} & 
\colhead{$K_s$}  & 
\colhead{$J-K_s$} \\
}
\startdata             
J0141+1804 & 01:41:03.20 +18:04:50.20 & L1 & L4.5 & 13.88$\pm$0.03 & 12.49$\pm$0.03 & 1.38$\pm$0.04 & Red &  1\\
J0247-1631 & 02:47:49.90 -16:31:12.60 & \nodata & T2$\pm$1.5 & 17.19$\pm$0.18 & 15.62$\pm$0.19 & 1.57$\pm$0.27 & SpB &  2 \\
J0351+4810 & 03:51:04.37 +48:10:46.80 & \nodata & T1$\pm$1.5 & 16.47$\pm$0.13 & 14.99$\pm$0.12 & 1.47$\pm$0.18 & Blue &  2 \\
J0407+1546 & 04:07:07.52 +15:46:45.70 & L3.5 & \nodata & 15.48$\pm$0.06 & 13.56$\pm$0.04 & 1.92$\pm$0.07 & Red &  3 \\
J0805+4812 & 08:05:31.84 +48:12:33.00 & L4 & L9$\pm$1.5 & 14.73$\pm$0.04 & 13.44$\pm$0.04 & 1.29$\pm$0.06 & SpB &   4 \\
J0825+2115 & 08:25:19.60 +21:15:52.10 & L7.5 & L6 & 15.10$\pm$0.03 & 13.03$\pm$0.03 & 2.07$\pm$0.04 & Red &  5 \\
J0939-2448 & 09:39:35.48 -24:48:27.90 & \nodata & T8 & 15.98$\pm$0.11 & $>$16.6 & -0.62 $\pm$ 0.11 & SpB &  6 \\
J1106+2754 & 11:06:11.91 +27:54:21.50 & \nodata & T2.5 & 14.82$\pm$0.04 & 13.80$\pm$0.05 & 1.02$\pm$0.07 & SpB &  7 \\
J1315-2649 & 13:15:30.90 -26:49:51.30 & L5.5 & \nodata & 15.20$\pm$0.05 & 13.46$\pm$0.04 & 1.73$\pm$0.07 & SpB &  8 \\
J1416+1348 & 14:16:24.09 +13:48:26.70 & L6 & L6p$\pm$2 & 13.15$\pm$0.03 & 12.11$\pm$0.02 & 1.03$\pm$0.03 & Blue &  9  \\
J1511+0607 & 15:11:14.66 +06:07:42.90 & \nodata & T0$\pm$2 & 16.02$\pm$0.08 & 14.54$\pm$0.10 & 1.47$\pm$0.13 & SpB &  2 \\
J1711+2232 & 17:11:45.70 +22:32:04.40 & L6.5 & \nodata & 17.09$\pm$0.18 & 14.73$\pm$0.10 & 2.36$\pm$0.20 & SpB &  5 \\
J2139+0220 & 21:39:26.76 +02:20:22.60 & \nodata & T1.5 & 15.26$\pm$0.05 & 13.58$\pm$0.05 & 1.68$\pm$0.07 & SpB &  3 \\
J2148+4003 & 21:48:16.28 +40:03:59.30 & L6 & L6p & 14.15$\pm$0.03 & 11.77$\pm$0.02 & 2.38$\pm$0.04 & Red &  10 \\
J2224-0158 & 22:24:43.80 -01:58:52.10 & L4.5 & L3.5 & 14.07$\pm$0.03 & 12.02$\pm$0.02 & 2.05$\pm$0.04 & Red &  5 \\
\enddata
\tablenotetext{a}{Red = unusually red; Blue = unusually blue; SpB = spectral binary.}
\tablerefs{(1) \citet{2003IAUS..211..197W} (2) \citet{2006AJ....131.2722C} (3) \citet{2008AJ....136.1290R} (4) \citet{2002AJ....123.3409H} (5) \citet{2000AJ....120..447K} (6) \citet{2005AJ....130.2326T} (7) \citet{2007AJ....134.1162L} (8) \citet{2002ApJ...564L..89H} (9) \citet{2010ApJ...710...45B} (10) \citet{2008ApJ...686..528L}.}
\end{deluxetable}

\clearpage

\begin{deluxetable}{c c c c c c}
\tablecaption{List of sources and observation information \label{observationtable}}
\tabletypesize{\footnotesize}
\tablewidth{0pt}
\tablehead{
\colhead{Source} &
\colhead{Obs.} &
\colhead{No. of} &
\colhead{Time} &
\colhead{Avg Seeing} &
\colhead{Airmass} \\
& \colhead{Date (UT)} & \colhead{dithers} & \colhead{(hrs)}& \colhead{($\arcsec$)} & \\
}
\startdata
J0805+4812 & 24-Mar-2010 & 25 & 1.991 & 1.61 & 1.02-1.05 \\
 & 28-Mar-2010 & 22 & 1.995 & 1.44 & 1.02-1.06 \\
\hline           
J0825+2115 & 23-Mar-2010 & 3 & 0.25 & 4.04 & 1.08-1.1 \\
 & 27-Mar-2010 & 6 & 0.586 & 1.97 & 1.15-1.2 \\
\hline
J0939-2448 & 28-Mar-2010 & 20 & 1.991 & 1.42 & 2.17-3.1 \\
\hline           
J1106+2754 & 23-Mar-2010 & 18 & 1.752 & 3.66 & 1.03-1.2 \\
 & 29-Mar-2010 & 21 & 1.895 & 1.69 & 1.04 \\
\hline
J1315-2649 & 24-Mar-2010 & 6 & 0.552 & 1.73 & 2.3-2.4 \\
 & 27-Mar-2010 & 6 & 0.528 & 1.96 & 2.35-2.51 \\
 & 28-Mar-2010 & 8 & 0.709 & 1.89 & 2.29-2.33 \\
 & 29-Mar-2010 & 6 & 0.551 & 2.04 & 2.29 \\
\hline                 
J1416+1348 & 23-Mar-2010 & 4 & 0.369 & 2.27 & 1.11-1.13 \\
 & 24-Mar-2010 & 6 & 0.562 & 1.36 & 1.13-1.17 \\
 & 27-Mar-2010 & 6 & 0.52 & 1.39 & 1.1-1.12 \\
 & 28-Mar-2010 & 6 & 0.518 & 2.31 & 1.12-1.15 \\
 & 29-Mar-2010 & 6 & 0.526 & 1.61 & 1.19 \\
\hline
J1511+0607 & 27-Mar-2010 & 6 & 0.523 & 1.66 & 1.29-1.37 \\
\hline           
J1711+2232 & 23-Mar-2010 & 8 & 0.683 & 3.21 & 1.03-1.04 \\
\hline                                       
J2139+0220 & 27-Oct-2010 & 3 & 1.603 & 2.57 & 1.25-1.9 \\
 & 28-Oct-2010 & 8 & 2.671 & 1.7 & 1.24-1.72 \\
 & 1-Nov-2010 & 3 & 0.216 & 1.77 & 1.22-1.25 \\
\hline           
J2148+4003 & 28-Oct-2010 & 3 & 1.227 & 1.52 & 1.01-1.12 \\
 & 1-Nov-2010 & 3 & 1.248 & 1.51 & 1-1.2 \\
\hline
J2224-0158 & 27-Oct-2010 & 9 & 2.565 & 2.87 & 1.29-1.77 \\
 & 28-Oct-2010 & 4 & 2.449 & 1.96 & 1.3-1.52 \\
 & 1-Nov-2010 & 3 & 1.217 & 1.6 & 1.29-1.37 \\
\hline
\enddata                
\end{deluxetable}

\clearpage

\begin{deluxetable}{ccc|cccccc}
\tablecaption{Results - J band\label{resultstable}}
\tabletypesize{\scriptsize}
\tablewidth{0pt}
\tablehead{
\colhead{Source} &
\colhead{Obs.} &
\colhead{No. of} &
\multicolumn{6}{c}{J-band} \\
& \colhead{date.} & \colhead{data pts} &  \colhead{$A_{var}$} &  \colhead{$\sigma_{med}$} &  \colhead{$S$} & \colhead{$\chi^2$} & \colhead{N$_{comp}$} &  \colhead{$\chi^2_{comp}$} \\
}
\startdata
J0805+4812 & 24-Mar-2010 &25&0.020&0.016&1.25&0.40&3&0.189\\
 & 28-Mar-2010 &22&0.034&0.022&1.55&0.78&4&0.474\\
&Combined&47&0.031&0.015&2.05&1.08&2&0.737\\
J0825+2115 & 23-Mar-2010 &3&0.015&0.015&1.00&1.30&3&0.246\\
 & 27-Mar-2010 &6&0.016&0.018&0.86&0.41&3&0.470\\
&Combined&9&0.033&0.022&1.51&0.95&3&0.668\\
J0939-2448 & 28-Mar-2010 &20&0.141&0.052&2.71&0.78&3&0.084\\
J1106+2754 & 23-Mar-2010 &18&0.022&0.014&1.57&0.88&3&0.516\\
 & 29-Mar-2010 &21&0.022&0.015&1.47&0.70&3&1.678\\
&Combined&39&0.039&0.014&2.87&1.21&3&0.841\\
 J1315-2649 & 24-Mar-2010 &6&0.027&0.021&1.29&1.00&5&0.448\\
 & 27-Mar-2010 &6&0.011&0.019&0.58&0.19&7&0.429\\
 & 28-Mar-2010 &8&0.016&0.020&0.80&0.45&5&0.549\\
 & 29-Mar-2010 &6&0.095&0.059&1.62& 2.07&3&0.797\\
 & Combined (24-28)&20&0.035&0.022&1.56&0.43&4&0.898\\
\textbf{J1416+1348} & 23-Mar-2010 &4&0.006&0.016&0.38&0.11&4&0.648\\
 & 24-Mar-2010 &6&0.069&0.114&0.61&0.24&11&0.378\\
 & 27-Mar-2010 &6&0.033&0.073&0.45&0.09&10&0.461\\
 & 28-Mar-2010 &6&0.028&0.025&1.12&0.78&2&0.738\\
 & \textbf{29-Mar-2010} &6&0.054&0.023&\textbf{2.35}&\textbf{2.67}&2&0.808\\
 &Combined&28&0.052&0.020&2.54&1.03&4&1.097\\
J1511+0607 & 27-Mar-2010 &6&0.036&0.047&0.77&0.30&10&0.384\\
\textbf{J1711+2232} & \textbf{23-Mar-2010} &7&0.103&0.041&\textbf{2.51}&\textbf{1.81}&5&0.387\\
\textbf{J2139+0220} & \textbf{27-Oct-2010} &8&0.062&0.017&\textbf{3.65}&\textbf{8.72}&3&1.412\\
 & 28-Oct-2010 &3&0.067&0.034&1.97&3.80&3&0.883\\
 & \textbf{1-Nov-2010} &3&0.038&0.017&\textbf{2.24}&\textbf{4.98}&3&0.520\\
 &\textbf{Combined}&14&0.067&0.020&\textbf{3.35}&\textbf{4.80}&3&0.999\\
J2148+4003 & 28-Oct-2010 &3&0.042&0.038&1.11&0.94&7&0.382\\
 & 1-Nov-2010 &3&0.010&0.019&0.53&0.29&14&0.274\\
J2224-0158 & 27-Oct-2010 &8&0.020&0.011&1.82&1.10&2&2.388\\
 & 28-Oct-2010 &3&0.044&0.020&2.20&6.04&1&5.974\\
 & 1-Nov-2010 &2&0.011&0.015&0.73&1.12&3&0.234\\ 
  \enddata
\end{deluxetable}

\begin{deluxetable}{ccc|cccccc}
\tablecaption{Results - K band\label{resultstablek}}
\tabletypesize{\scriptsize}
\tablewidth{0pt}
\tablehead{
\colhead{Source} &
\colhead{Obs.} &
\colhead{No. of} &
\multicolumn{6}{c}{K-band} \\
& \colhead{date.} & \colhead{data pts} &  \colhead{$A_{var}$} &  \colhead{$\sigma_{med}$} &  \colhead{$S$} & \colhead{$\chi^2$} & \colhead{N$_{comp}$} &  \colhead{$\chi^2_{comp}$} \\
}
\startdata
J0805+4812 & 24-Mar-2010 &25&0.026&0.019&1.37&0.60&3&0.80\\
 & 28-Mar-2010 &22&0.083&0.038&2.18&1.24&4&0.74\\
&Combined&47& 0.068& 0.019& 3.52& 1.93& 2& 1.60\\
J0825+2115 & 23-Mar-2010 &3&0.006&0.015&0.40&0.18&3&0.06\\
 & 27-Mar-2010 &6&0.020&0.026&0.77&0.29&4&0.38\\
&Combined&9& 0.012& 0.013& 0.90& 0.30& 3& 0.45\\
\textbf{J0939-2448} & \textbf{28-Mar-2010} &20&0.310&0.094&\textbf{3.30}&\textbf{2.51}&3&0.73\\
J1106+2754 & 23-Mar-2010 &18&0.031&0.021&1.48&0.79&3&0.93\\
 & 29-Mar-2010 &21&0.047&0.021&2.24&1.18&3&1.06\\
&Combined&39&0.038& 0.02&1.9&0.75& 3& 0.96\\
J1315-2649 & 24-Mar-2010 &6&0.025&0.028&0.89&0.50&7&0.41\\
 & 27-Mar-2010 &6&0.020&0.021&0.95&0.70&6&0.40\\
 & 28-Mar-2010 &8&0.100&0.021&4.76&6.10&2&3.37\\
 & 29-Mar-2010 &6&0.158&0.028&5.64&12.1&2&8.71\\
 & Combined (24-28)&20& 0.078& 0.019& 4.07& 2.58& 3& 2.29\\
J1416+1348 & 24-Mar-2010 &6&0.045&0.036&1.25&0.66&2&0.85\\
 & 27-Mar-2010 &6&0.022&0.025&0.88&0.43&2&0.94\\
 & 28-Mar-2010 &6&0.024&0.046&0.52&0.25&3&1.24\\
 & 29-Mar-2010 &6&0.045&0.033&1.36&0.73&2&1.26\\
 &Combined&24& 0.052& 0.034& 1.55& 0.59& 2& 1.17\\
J1511+0607 & 27-Mar-2010 &6&0.061&0.036&1.69&1.84&5&0.47\\
J1711+2232 & 23-Mar-2010 &7&0.593&0.083&7.14&18.3&4&3.39\\
J2139+0220 & 27-Oct-2010 &8&0.065&0.034&1.91&1.91&2&3.65\\
 & 28-Oct-2010 &3&0.035&0.068&0.51&0.26&2&1.23\\
 & 1-Nov-2010 &3&0.031&0.044&0.70&0.89&2&1.01\\
 &Combined&14& 0.08& 0.04& 2.02& 1.21& 2& 2.09\\
J2224-0158 & 27-Oct-2010 &8&0.063&0.025&2.52&2.66&1&3.33\\
 & 28-Oct-2010 &3&0.105&0.022&4.77&15.8&2&5.09\\
 & 1-Nov-2010 &2&0.012&0.017&0.71&1.07&3&0.09\\ 
   \enddata
\end{deluxetable}

\begin{deluxetable}{ccc}
\tablecaption{Radial Velocity Measurements for J2139+0220  \label{tab:rv}}
\tabletypesize{\scriptsize}
\tablewidth{0pt}
\tablehead{
\colhead{Date} & 
\colhead{MJD\tablenotemark{a}} & 
\colhead{RV} \\
\colhead{(UT)} & & \colhead{({\kms})} \\ 
}
\startdata             
2010 Nov 26 & 55526.197 & -23.3$\pm$1.1 \\
2011 Jul 6 & 55748.543 & -26.9$\pm$1.5 \\
2011 Sep 7 & 55811.377 & -27.5$\pm$1.2 \\
\enddata
\tablenotetext{a}{Modified Julian Date: Julian Date - 2450000}
\end{deluxetable}

\clearpage

\begin{deluxetable}{ccccccc}
\tablecaption{Variability fraction of L and T dwarfs for distinct spectral ranges  \label{tab:var2}}
\tabletypesize{\scriptsize}
\tablewidth{0pt}
\tablehead{
\colhead{Paper} & 
\colhead{SpType} & 
\colhead{Total} &
\colhead{Number} &
\colhead{Fraction} &
\colhead{Detection Limit} &
\colhead{Filters}\\
 & \colhead{Range} & \colhead{Observed}&\colhead{Variable} & & (mags) & \\ 
}
\startdata
Bailer-Jones and Mundt (2001) & L0-L5 & 10 & 7 & $70^{+10}_{-17}\%$ &0.01& I \\             
\\[1pt]
\citet{2002ApJ...577..433G} & L0-L4.5 & 18 & 7 & $39^{+12}_{-10}\%$ &0.01 & Ic\\
\\[1pt]
\citet{2003MNRAS.346..473K} & L0-T2.5 & 12 & 3 & $25^{+16}_{-8}\%$ &0.016 & Ic\\
\\[1pt]
\citet{2003AJ....126.1006E} & L2-T5 & 9 & 3 & $33^{+18}_{-11}\%$ &0.36 & Ks\\
\\[1pt]
\citet{2004MNRAS.354..466K} & L0-T8 & 18 & 6 & $33^{+12}_{-9}\%$ &0.02& J,H,K\\
\\[1pt]
\citet{2005MNRAS.360.1132K} & L1-T6 & 12 & 3 & $25^{+16}_{-8}\%$ &0.02& Ic\\
\\[1pt]
Khandrika et. al (2012) & L3.5-T8 & 15 & 4 & $27^{+14}_{-8}\%$ &0.05& J,K'\\
\\
\hline
\\
\textbf{Total }&\textbf{L0-L5} & \textbf{49 \tablenotemark{a} } &\textbf{18 \tablenotemark{a} } & $\mathbf{36^{+7}_{-6} }\%$ & & \\
\\
\textbf{Total }&\textbf{L6-T0} & \textbf{13 \tablenotemark{a} } &\textbf{4 \tablenotemark{a} } & $\mathbf{30^{+14}_{-9} }\%$ & & \\
\\
\textbf{Total }&\textbf{T1-T5} & \textbf{11 \tablenotemark{a} } &\textbf{4 \tablenotemark{a} } & $\mathbf{36^{+15}_{-11} }\%$ & & \\
\\
\textbf{Total }&\textbf{T6-T8} & \textbf{5 \tablenotemark{a} } &\textbf{1 \tablenotemark{a} } & $\mathbf{20^{+25}_{-7} }\%$ & & \\
\\
\enddata
\tablenotetext{a}{The total value is less than the sum as it contains no repeatedly-observed sources}
\end{deluxetable}

\clearpage


\end{document}